\documentclass[aps]{revtex4}
\usepackage{amsfonts}
\usepackage{amsmath}
\usepackage{amssymb,epsf}

\begin{document}

\title{Massive charged BTZ black holes in asymptotically (a)dS spacetimes}
\author{S. H. Hendi$^{1,2}$\footnote{
email address: hendi@shirazu.ac.ir}, B. Eslam Panah$^{1}$\footnote{%
email address: behzad.eslampanah@gmail.com}, and S.
Panahiyan$^{1,3}$\footnote{ email address:
sh.panahiyan@gmail.com}} \affiliation{$^1$ Physics Department and
Biruni Observatory, College of Sciences, Shiraz
University, Shiraz 71454, Iran\\
$^2$ Research Institute for Astronomy and Astrophysics of Maragha (RIAAM), P.O. Box 55134-441, Maragha, Iran\\
$^3$Physics Department, Shahid Beheshti University, Tehran 19839,
Iran}

\begin{abstract}
Motivated by recent developments of BTZ black holes and interesting results
of massive gravity, we investigate massive BTZ black holes in the presence
of Maxwell and Born-Infeld (BI) electrodynamics. We study geometrical
properties such as type of singularity and asymptotical behavior as well as
thermodynamic structure of the solutions through canonical ensemble. We show
that despite the existence of massive term, obtained solutions are
asymptotically (a)dS and have a curvature singularity at the origin. Then,
we regard varying cosmological constant and examine the Van der Waals like
behavior of the solutions in extended phase space. In addition, we employ
geometrical thermodynamic approaches and show that using Weinhold, Ruppeiner
and Quevedo metrics leads to existence of ensemble dependency while HPEM
metric yields consistent picture. For neutral solutions, it will be shown
that generalization to massive gravity leads to the presence of non-zero
temperature and heat capacity for vanishing horizon radius. Such behavior is
not observed for linearly charged solutions while generalization to
nonlinearly one recovers this property.
\end{abstract}

\maketitle

\section{Introduction}

Three dimensional black holes were first studied by Banados, Teitelboim and
Zanelli which are known as BTZ Black holes \cite{BTZ1,BTZ2,BTZ3}. The
discovery of the BTZ black holes contributed to studies that are conducted
in different aspects of physics. Among them one can point out: providing
simplified machinery for studying different properties of black holes such
as thermodynamical ones \cite%
{BTZthermo1,BTZthermo2,BTZthermo3,BTZthermo4,BTZthermo5,BTZthermo6,BTZthermo7}%
, contributing to our understanding of gravitational systems and their
interactions in lower dimensions \cite{BTZinter}, existence of specific
relations between BTZ black holes and effective action in string theory \cite%
{BTZstring1,BTZstring2,BTZstring3,BTZstring4} and possible existence of
gravitational Aharonov-Bohm effect due to the noncommutative BTZ black holes
\cite{BohmBTZ}. In addition, several studies were conducted in context of
AdS/CFT correspondence \cite{BTZAdS1,BTZAdS2,BTZAdS3}, quantum aspect of
three dimensional gravity, entanglement, quantum entropy \cite%
{BTZent1,BTZent2,BTZent3,BTZent4,BTZent5,BTZent6,BTZent7} and holographic
aspects of such solutions \cite{BTZholo1,BTZholo2,BTZholo3}. Furthermore,
three dimensional black holes and their thermodynamical properties in the
presence of different matter fields with various gravity models are
addressed in literature \cite{BTZZ1,BTZZ2,BTZZ3,BTZZ4,BTZZ5,BTZZ6}.
Motivated by the above brief discussion, we will conduct our study in three
dimensional spacetime.

On the other hand, Einstein theory of gravity introduces gravitons as
massless particles with two degrees of freedom, whereas there are several
arguments that state gravitons should be massive particles. In order to have
massive graviton, theory of general relativity should be modified to include
mass terms as well. The first attempt for constructing a massive theory was
referred to the works of Fierz and Pauli \cite{Fierz}. This theory was built
in a flat background and its generalization to curved background leads to
the existence of a typical ghost (Boulware-Deser ghost) \cite{Deser}.
Presence of the ghost indicates that the theory under consideration is
unstable. In order to avoid such instability, several models of massive
theory are proposed. One of the ghost-free massive theories was introduced
in a pioneering work by Bergshoeff, Hohm and Townsend. This theory is a
three dimensional massive theory and is known as new massive gravity (NMG)
\cite{NEW}. The construction of NMG was done based on two important facts:
first, a massless graviton has no propagating degrees of freedom in three
dimensions. This leads to a symmetric tensor in three dimensions with six
components with a diffeomorphism symmetry. Due to this diffeomorphism
symmetry, three degrees of freedom are pure gauge and the other three ones
are non-dynamical. Second, a massive graviton in three dimensions has two
degrees of freedom similar to a massless graviton in four dimensions.
Therefore, it is possible to build a diffeomorphism invariant theory of
massive gravity in which only the massive degree of freedom is present.
Black hole solutions in presence of this massive gravity theory have been
obtained in literature \cite%
{NewBlack1,NewBlack2,NewBlack3,NewBlack4,NewBlack5,NewBlack6,NewBlack7}. In
addition, some of their stabilities have been addressed in Refs. \cite%
{NewS1,NewS2}. As for other three dimensional massive gravities, one can
regard topological massive gravity \cite{topol}, supergravity extensions
\cite{super} and critical gravity \cite{crit} (for more details, see Ref.
\cite{review}).

Another class of massive gravity was proposed by de Rham, Gabadadze and
Tolley (dRGT) \cite{de1,de2}. Contrary to previous three dimensional
theories, dRGT theory is valid in higher dimensions as well. In this theory,
the mass terms are produced by consideration of a reference metric. The
stability of this theory was addressed in Refs. \cite{stabal1,stabal2} and
it was shown that such theory enjoys absence of the Boulware-Deser ghost
\cite{stabal1,stabal2}. Black hole solutions of dRGT massive gravity and
their thermodynamical properties have been investigated in Ref. \cite{dRGTb1}%
. In addition, the stability of the Schwarzschild-de Sitter black holes in
presence of dRGT massive gravity has been analyzed \cite{dRGTb2} (For
detailed study regarding static spherically symmetric black holes in dRGT
massive gravity, we refer the reader to Ref. \cite{dRGTb3}).

Motivated by applications of gauge/gravity duality, Vegh introduced another
type of massive gravity theory \cite{Vegh}. In this theory, similar to dRGT
theory, mass terms are built by using a reference metric except that this
reference metric is a singular one. Using this theory of massive gravity,
Vegh has shown that graviton may behave like a lattice and exhibits a Drude
peak \cite{Vegh}. In Ref. \cite{stabalVegh}, it was pointed out that for
arbitrary singular metric, this theory of massive gravity is ghost-free and
stable. Charged black hole solutions in the presence of this massive gravity
have been studied in Ref. \cite{CaiMassive}. Also, the existence of Van der
Waals like behavior in extended phase space has been addressed in Ref. \cite%
{extend}. Moreover, the generalizations of such theory to include nonlinear
electrodynamics, higher derivative gravity and gravity's rainbow have been
done \cite{HendiMassive1,HendiMassive2,HendiMassive3,HendiMassive4}. In
addition, thermodynamic properties, phase transitions and thermal stability
of such generalized models have been investigated in Refs. \cite%
{HendiMassive1,HendiMassive2,HendiMassive3,HendiMassive4}. Furthermore, it
was shown that the mentioned massive gravity has sensible effects in
metal/superconductor phase transition. It was pointed out that for this
phase transition, the maximal tunnelling current and the coherence length of
junction are affected by the mass of graviton in Vegh's massive gravity
model \cite{peng hu}. In addition, phase transition of holographic
entanglement entropy in this massive gravity is investigated as well \cite%
{entMassive}. It is also notable that the existence of massive graviton
results into new phenomenology for gravitational system. Among them one can
name: modification in propagation's speed of the gravitational waves \cite%
{GW1}, effects on production of gravitational wave during inflation \cite%
{GW2,GW3} and existence of additional polarization for gravitational waves.
The mentioned arguments motivate us to study black hole solutions in the
presence of Vegh's massive gravity.

In order to have a deep insight for considering three dimensional massive
gravity, one may regard an electromagnetic field as a source. Maxwell theory
of electrodynamics has been employed to study electrically charged systems
in various branches of physics. Although Maxwell theory is a relatively
successful theory in describing different phenomena, it is not flawless. One
of the main problems of this theory is existence of singularity for
electromagnetic field of point-like charges at their locations. In other
words, the electric field of a point-like charge diverges at the charge
location which leads to an infinite self-energy. In 1934, Born and Infeld
(BI) \cite{BI} introduced a nonlinear electromagnetic field theory to solve
infinite self-energy problem by imposing a maximum strength of the
electromagnetic field. One year later, this nonlinear theory was first
employed by Hoffmann in context of Einstein gravity \cite{Hoffmann}.
Although BI theory has long been discarded due to its complexity, in recent
two decades, different types of black holes in the presence of BI
electrodynamics have been investigated \cite%
{BIpapers1,BIpapers2,BIpapers3,BIpapers4,BIpapers5,BIpapers6,BIpapers7,BIpapers8,BIpapers9,BIpapers10,BIpapers11,BIpapers12}%
.

It is worthwhile to mention that the generalization of Maxwell field to
nonlinear electrodynamics leads to enriching the new solutions and
introducing a new phenomenology for the system under consideration. Among
various motivations for studying BI theory, one can point out obtaining BI
effective action in superstring theory and absence of singularity for
D-branes \cite{Fradkin1,Fradkin2,Fradkin3,Fradkin4,Fradkin5,Gibbons}, and
also Eddington-inspired BI theory's applications in gravitation and
cosmology \cite%
{Eddington1,Eddington2,Eddington3,Eddington4,Eddington5,Eddington6,Eddington7}%
. In addition, the functional form of BI nonlinear theory of electrodynamics
has been employed as gravitational theory as well \cite{Born-G1,Born-G2}.
For duality and other aspects of BI theory, we refer the reader to some
interesting papers \cite%
{Dulaization1,Dulaization2,Dulaization3,Dulaization4,Dulaization5,Dulaization6,Dulaization7}%
. Due to these motivations and in order to have a better picture of charged
three dimensional black holes in presence of massive gravity, we study two
cases of linearly (Maxwell theory) and nonlinearly (BI theory) charged black
hole solutions and investigate their thermodynamical properties.

Thermodynamical aspect of black holes has been of a great interest ever
since the pioneering work of Hawking and Beckenstein \cite%
{hawking1,hawking2,hawking3}. Recent progresses in gauge/gravity duality has
also highlighted importance of black hole thermodynamics \cite%
{1AdSCFT1,AdSCFT1,AdSCFT2,AdSCFT3,AdSCFT4,AdSCFT5,AdSCFT6,AdSCFT7,AdSCFT8,AdSCFT9,AdSCFT10,AdSCFT11,2AdSCFT1,3AdSCFT1,4AdSCFT1}%
. In addition, the black hole thermodynamics plays significant role in
non-perturbative aspects of quantum gravity. In other words, thermodynamical
aspect of black holes may provide a basis for constructing a consistent
theory of quantum gravity. Considering the mentioned motivations, we study
some thermodynamical properties of three dimensional charged massive black
holes. Among the thermodynamical properties, thermal stability has specific
importance \cite{Myung1,Myung2,Myung3,Myung4}. In order to black holes being
physical thermodynamical systems, they should be thermally stable. In case
of unstable solution, system may go under phase transition to acquire stable
state. In this paper, we also study thermal stability and phase transition
of black holes through canonical ensemble.

Another method of studying thermodynamical structure of black holes is
through the use of geometrical method. In other words, by considering a
thermodynamical quantity as a potential and its corresponding extensive
parameters, one can build a phase space of the thermodynamical system. The
Ricci scalar of this phase space carries information regarding
thermodynamical properties. The divergencies of thermodynamical Ricci scalar
may coincide to bound or phase transition points. So far, four different
geometrical methods were introduced; Weinhold \cite{WeinholdI,WeinholdII},
Ruppeiner \cite{RuppeinerI,RuppeinerII}, Quevedo \cite{QuevedoI,QuevedoII}
and HPEM \cite{HPEMI,HPEMII,HPEMIII}. Several studies in context of black
holes with consideration of such methods have been done \cite%
{HanC,BravettiMMA,Ma,GarciaMC,ZhangCY,MoLW2016}. In addition, it was pointed
out that one can employ geometrical thermodynamics to obtain phase
transition points in superconductors \cite{BasakCNS}. Furthermore, a
comparative study in context of black holes was done in Ref. \cite{comp}.
Thermodynamical concepts indicate that studying a thermodynamical system
should lead to same results irrespective of employed ensemble. In other
words, no ensemble dependency must exist. The existence of ensemble
dependency in geometrical thermodynamics \cite{dependency} could be observed
through two cases: I) bound and/or phase transition points is/are not
matched with divergency of thermodynamical Ricci scalar. II) extra
divergency exists for the Ricci scalar which is not matched with any
mentioned points. In several studies, it was pointed out that employing
Weinhold, Ruppeiner and Quevedo metrics leads to existence of ensemble
dependency \cite{HPEMI,HPEMII,HPEMIII}. In order to avoid such problem, HPEM
metric was proposed. Motivated by the applications of geometrical
thermodynamics and possible existence of ensemble dependency, in this paper,
we conduct a comparative study regarding different approaches of geometrical
thermodynamics for three dimensional charged black holes in massive gravity.
Linearly and nonlinearly charged three dimensional black hole solutions and
their thermodynamical properties were investigated in Refs. \cite%
{HPEMI,HendiJHEP,Mamasani}. In this paper, we generalize our solutions to
include massive gravity. We examine the effects of this generalization on
thermodynamical aspect of charged three dimensional black holes.

The outline of the paper will be as follow. In Sec. \ref{FieldEq}, we
introduce action and basic field equations related to three dimensional
charged black holes in the massive gravity context. Section \ref%
{Einstein-Maxwell} is devoted to obtain black hole solutions of the massive
gravity in linearly charged set up. The thermodynamics and thermal stability
will be investigated and some remarks regarding neutral case are given.
Next, generalization to nonlinear electrodynamics is done and the effects of
this generalization on thermodynamics and phase transition of the black
holes are investigated. In addition, it will be shown that these black
holes, despite generalization of the massive gravity and nonlinear
electromagnetic fields, suffer the absence of Van der Waals like behavior in
their phase spaces. Furthermore, geometrical thermodynamics is employed to
study phase transition of these black holes. The paper is concluded by some
closing remarks.

\section{Basic Equations}

\label{FieldEq}

Here, we start with the $3$-dimensional action of Einstein-massive gravity
with an abelian $U(1)$ gauge field
\begin{equation}
\mathcal{I}=-\frac{1}{16\pi }\int d^{3}x\sqrt{-g}\left[ \mathcal{R}-2\Lambda
+L(\mathcal{F})+m^{2}\sum_{i}^{4}c_{i}\mathcal{U}_{i}(g,f)\right] ,
\label{Action}
\end{equation}%
where $\mathcal{R}$ and $L(\mathcal{F})$ are, respectively, the scalar
curvature and an arbitrary Lagrangian of electrodynamics, $\Lambda$ is the
cosmological constant and $f$ is a fixed symmetric tensor. Also, $\mathcal{F}%
=F_{\mu \nu }F^{\mu \nu }$\ is the Maxwell invariant, in which $F_{\mu \nu
}=\partial _{\mu }A_{\nu }-\partial _{\nu }A_{\mu }$ is the Faraday tensor
and $A_{\mu } $ is the gauge potential. In addition, $c_{i}$'s are some
constants and $\mathcal{U}_{i}$'s are symmetric polynomials of the
eigenvalues of the $d\times d$ matrix $\mathcal{K}_{\nu }^{\mu }=\sqrt{%
g^{\mu \alpha }f_{\alpha \nu }}$ which can be written as follows
\begin{eqnarray}
\mathcal{U}_{1} &=&\left[ \mathcal{K}\right] ,\;\;\;\;\;\mathcal{U}_{2}=%
\left[ \mathcal{K}\right] ^{2}-\left[ \mathcal{K}^{2}\right] ,\;\;\;\;\;%
\mathcal{U}_{3}=\left[ \mathcal{K}\right] ^{3}-3\left[ \mathcal{K}\right] %
\left[ \mathcal{K}^{2}\right] +2\left[ \mathcal{K}^{3}\right] ,  \notag \\
&&\mathcal{U}_{4}=\left[ \mathcal{K}\right] ^{4}-6\left[ \mathcal{K}^{2}%
\right] \left[ \mathcal{K}\right] ^{2}+8\left[ \mathcal{K}^{3}\right] \left[
\mathcal{K}\right] +3\left[ \mathcal{K}^{2}\right] ^{2}-6\left[ \mathcal{K}%
^{4}\right] .  \notag
\end{eqnarray}

Taking into account the action (\ref{Action}) and using the variational
principle, we can obtain the field equations related to the gravitation and
gauge fields as
\begin{equation}
G_{\mu \nu }+\Lambda g_{\mu \nu }-\frac{1}{2}g_{\mu \nu }L(\mathcal{F})-2L_{%
\mathcal{F}}F_{\mu \rho }F_{\nu }^{\rho }+m^{2}\chi _{\mu \nu }=0,
\label{Field equation}
\end{equation}%
\begin{equation}
\partial _{\mu }\left( \sqrt{-g}L_{\mathcal{F}}F^{\mu \nu }\right) =0,
\label{Maxwell equation}
\end{equation}%
where $G_{\mu \nu }=R_{\mu \nu }-\frac{1}{2}g_{\mu \nu }R$, $L_{\mathcal{F}}=%
\frac{dL(\mathcal{F})}{d\mathcal{F}}$ and $\chi _{\mu \nu }$ is the massive
term with the following form
\begin{eqnarray}
\chi _{\mu \nu } &=&-\frac{c_{1}}{2}\left( \mathcal{U}_{1}g_{\mu \nu }-%
\mathcal{K}_{\mu \nu }\right) -\frac{c_{2}}{2}\left( \mathcal{U}_{2}g_{\mu
\nu }-2\mathcal{U}_{1}\mathcal{K}_{\mu \nu }+2\mathcal{K}_{\mu \nu
}^{2}\right) -\frac{c_{3}}{2}(\mathcal{U}_{3}g_{\mu \nu }-3\mathcal{U}_{2}%
\mathcal{K}_{\mu \nu }+  \notag \\
&&6\mathcal{U}_{1}\mathcal{K}_{\mu \nu }^{2}-6\mathcal{K}_{\mu \nu }^{3})-%
\frac{c_{4}}{2}(\mathcal{U}_{4}g_{\mu \nu }-4\mathcal{U}_{3}\mathcal{K}_{\mu
\nu }+12\mathcal{U}_{2}\mathcal{K}_{\mu \nu }^{2}-24\mathcal{U}_{1}\mathcal{K%
}_{\mu \nu }^{3}+24\mathcal{K}_{\mu \nu }^{4}).  \label{massiveTerm}
\end{eqnarray}

\section{Einstein-Maxwell solutions in the context of massive gravity:}

\label{Einstein-Maxwell}

In this section, we regard linearly charged three dimensional solutions in
the context of massive gravity and investigate their geometric and
thermodynamic properties.

\subsection{Linearly charged black hole solutions}

Here, we obtain static charged black holes with (a)dS asymptotes. For this
purpose, we consider the metric of $3$-dimensional spacetime with $(- + +)$
signature in the following explicit form
\begin{equation}
ds^{2}=-f(r)dt^{2}+f^{-1}(r)dr^{2}+r^{2}d\varphi ^{2},  \label{metric}
\end{equation}
where $f(r)$ is an arbitrary function of radial coordinate.

In order to obtain exact solutions, we should make a choice for the
reference metric. We consider the following ansatz metric \cite{CaiMassive}
\begin{equation}
f_{\mu \nu }=diag(0,0,c^{2}h_{ij}),  \label{f11}
\end{equation}%
where $c$ is a positive constant. Using the metric ansatz (\ref{f11}), $%
\mathcal{U}_{i}$'s are in the following forms \cite{CaiMassive}
\begin{equation}
\mathcal{U}_{1}=\frac{c}{r},\;\;\;\;\;\mathcal{U}_{2}=\mathcal{U}_{3}=%
\mathcal{U}_{4}=0,  \label{U}
\end{equation}
where indicate that since we are working in three dimensions, the only
contribution of massive gravity comes from $\mathcal{U}_{1}$.

Due to the fact that we are going to study the linearly charged BTZ
solutions, we choose the Lagrangian of Maxwell field $L(\mathcal{F})=-%
\mathcal{F}$. In addition, due to our interest in electrically charged black
holes in massive gravity context, we consider a radial electric field which
its related gauge potential is
\begin{equation}
A_{\mu }=h\left( r\right) \delta _{\mu }^{t}.  \label{gauge
potential}
\end{equation}

Using the metric (Eq. (\ref{metric})) with the Maxwell field equation (Eq. (%
\ref{Maxwell equation})), one finds the following differential equation
\begin{equation}
h^{\prime}(r)+rh^{\prime \prime }(r)=0,  \label{heq}
\end{equation}
where the prime and double prime are the first and the second derivatives
versus $r$, respectively. It is easy to find the solution of Eq. (\ref{heq})
as
\begin{equation}
h(r)=q\ln \left( \frac{r}{l}\right) ,  \label{h(r)}
\end{equation}%
where $q$ is an integration constant which is related to the electric charge
and $l$ is an arbitrary constant with length dimension which is coming from
the fact that the logarithmic arguments should be dimensionless. It is
notable that the electromagnetic field tensor is $F_{tr}=\frac{q}{r}$, which
does not depend on $l$ as a measurable physical quantity.

Now, we want to obtain exact solutions for the metric function $f(r)$. For
this purpose, by employing Eq. (\ref{metric}) with Eq. (\ref{Field equation}%
), we obtain the following differential equations
\begin{eqnarray}
&&rf^{\prime }(r)+2r^{2}\Lambda +2q^{2}-m^{2}cc_{1}r =0,  \label{eqENMax1} \\
&&\frac{r^{2}}{2}f^{\prime \prime }(r)+\Lambda r^{2}-q^{2} =0,
\label{eqENMax2}
\end{eqnarray}%
which are corresponding to $tt $(or $rr$) and $\varphi \varphi $ components
of Eq. (\ref{Field equation}), respectively. After some manipulations, one
can obtain the following metric function
\begin{equation}
f(r)=-\Lambda r^{2}-m_{0}-2q^{2}\ln \left( \frac{r}{l}\right) +m^{2}cc_{1}r,
\label{f(r)ENMax}
\end{equation}%
where $m_{0}$ is an integration constant which is related to the total mass
of the black hole. We should note that the obtained metric function
satisfies all components of the field equation (\ref{Field equation}),
simultaneously. Also, it is notable that, in the absence of massive
parameter ($m=0$), Eq. (\ref{f(r)ENMax}) reduces to the following linearly
charged BTZ black hole
\begin{equation}
f\left( r\right) =-\Lambda r^{2}-m_{0}-2q^{2}\ln \left( \frac{r}{l}\right) .
\end{equation}

Now, we are in a position to examine the geometrical structure of this
solution, in brief. First, we look for essential singularity(ies). The Ricci
and Kretschmann scalars are
\begin{eqnarray}
R&=&6\Lambda +\frac{2q^{2}}{r^{2}}-\frac{2m^{2}cc_{1}}{r}, \\
R_{\alpha \beta \gamma \delta }R^{\alpha \beta \gamma \delta } &=&12\Lambda
^{2}-\frac{8m^{2}cc_{1}\Lambda }{r}+\frac{2\left(
m^{4}c^{2}c_{1}^{2}+4q^{2}\Lambda \right) }{r^{2}}-\frac{8q^{2}m^{2}cc_{1}}{%
r^{3}}+\frac{12q^{4}}{r^{4}},
\end{eqnarray}%
which diverge at the origin and confirm that there is a curvature
singularity at $r=0$. For large values of radial coordinate, $%
r\longrightarrow \infty $, the Ricci and Kretschmann scalars are,
respectively, $6\Lambda $ and $12\Lambda ^{2}$, in which confirm that the
asymptotical behavior of the solution is (a)dS for $\Lambda >0$ ($\Lambda <0$%
).

\begin{figure}[tbp]
$%
\begin{array}{ccc}
\epsfxsize=5cm \epsffile{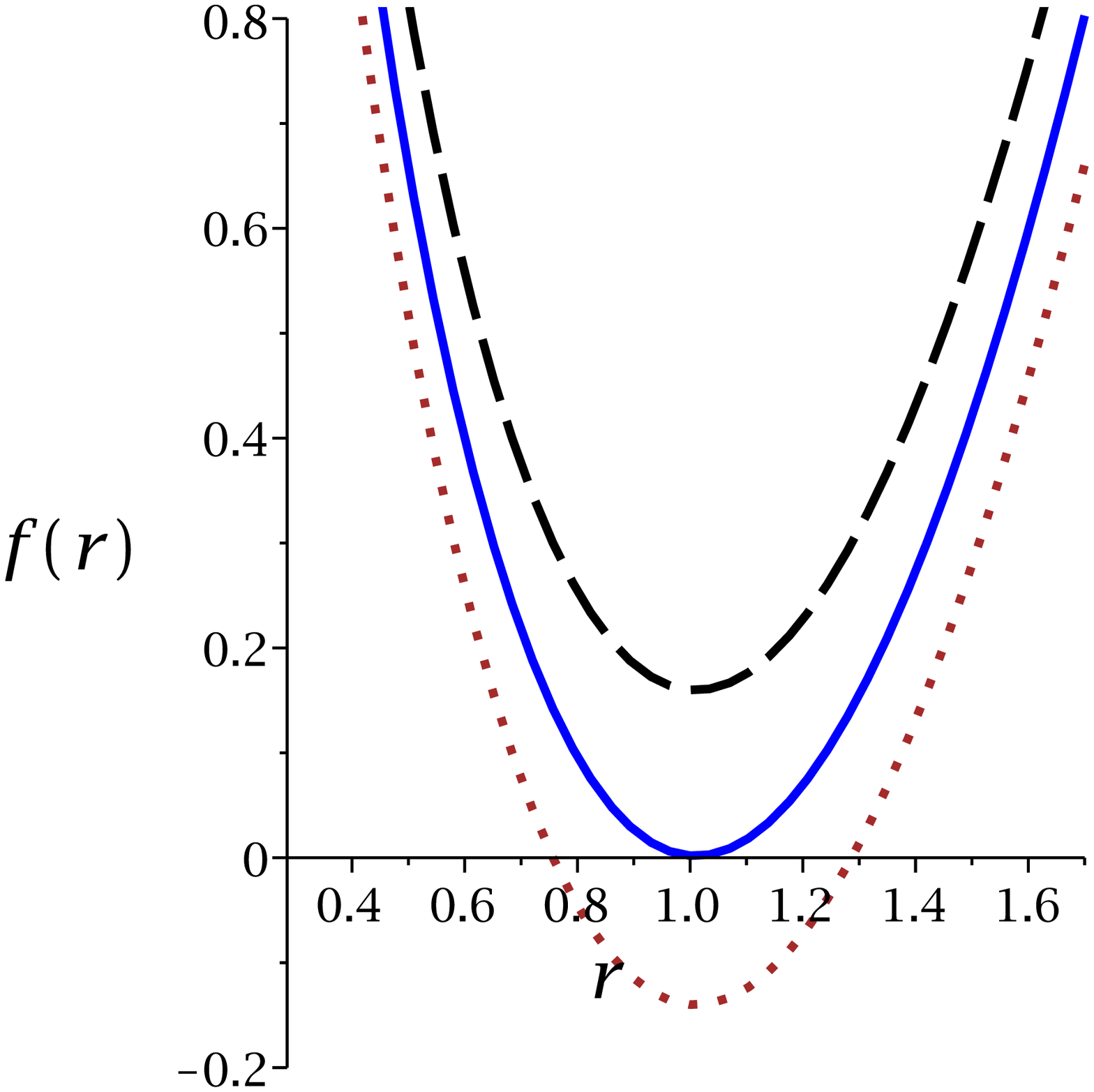} & \epsfxsize=5cm %
\epsffile{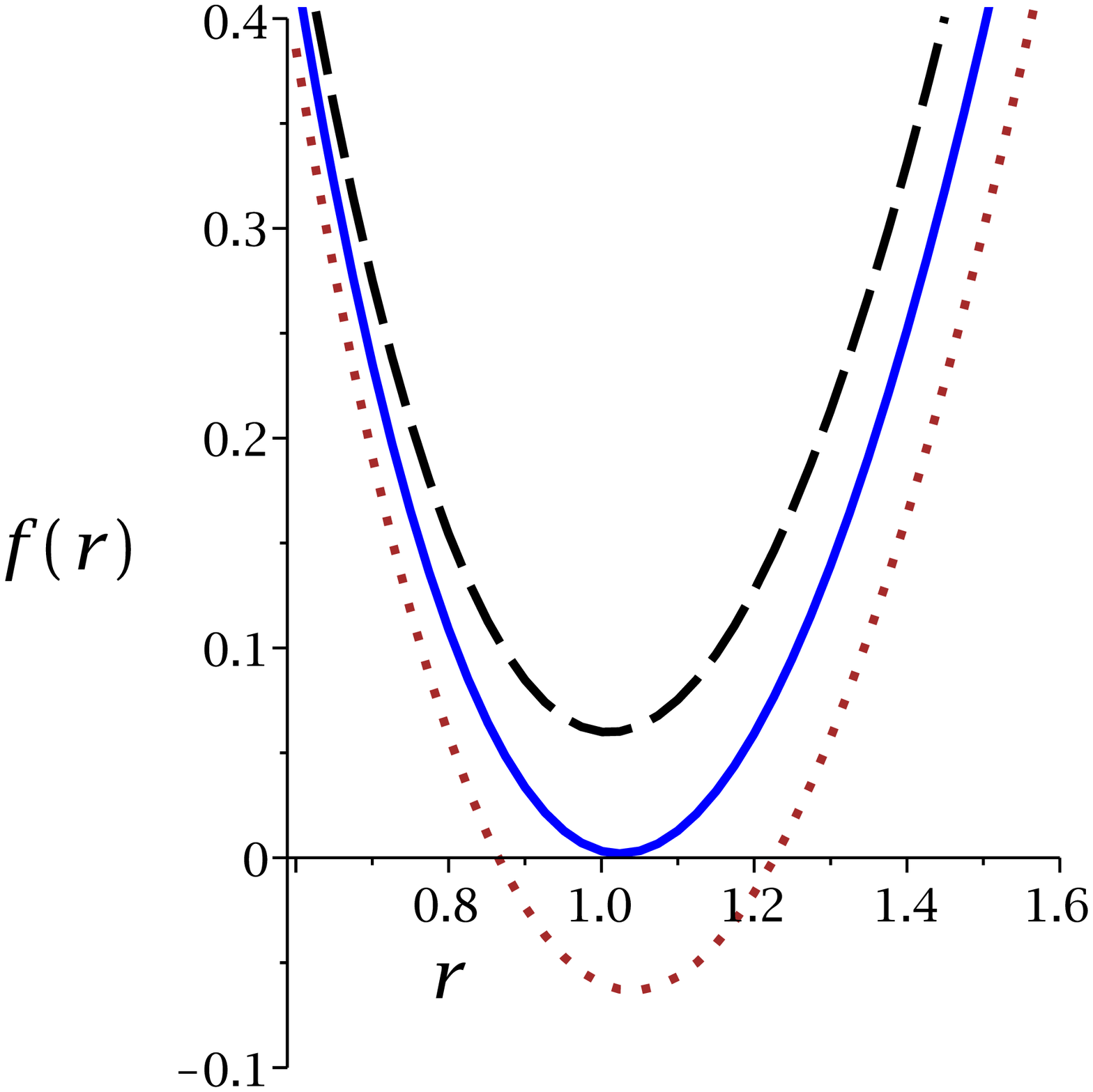} & \epsfxsize=5cm \epsffile{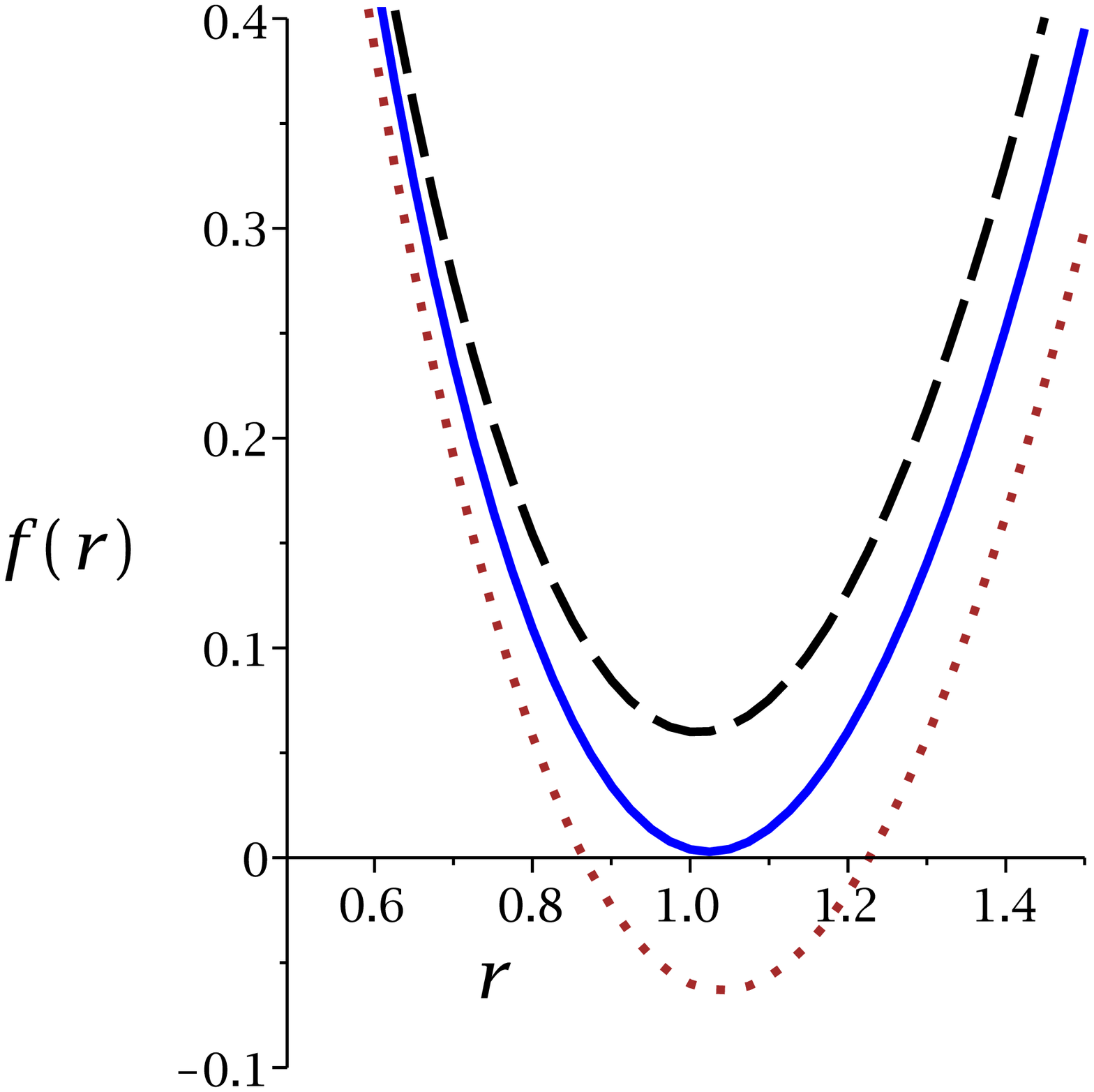}%
\end{array}
$%
\caption{\textbf{Eq. (\protect\ref{f(r)ENMax}):} $f(r)$ versus $r$ for $%
\Lambda=-1$, $q=1$, $m=0.2$ and $l=1$. \newline
Left panel: $c=1$, $c_{1}=-1$, $m_{0}=0.8$ (dashed line), $m_{0}=0.96$
(continuous line) and $m_{0}=1.1$ (dotted line). \newline
Middle panel: $c=1$, $m_{0}=0.9$, $c_{1}=-1$ (dashed line), $c_{1}=-2.42$
(continuous line) and $c_{1}=-4$ (dotted line). \newline
Right panel: $c_{1}=-1$, $m_{0}=0.9$, $c=1$ dashed line), $c=2.4$
(continuous line) and $c=4$ (dotted line).}
\label{Figfr1}
\end{figure}

\begin{figure}[tbp]
$%
\begin{array}{ccc}
\epsfxsize=5cm \epsffile{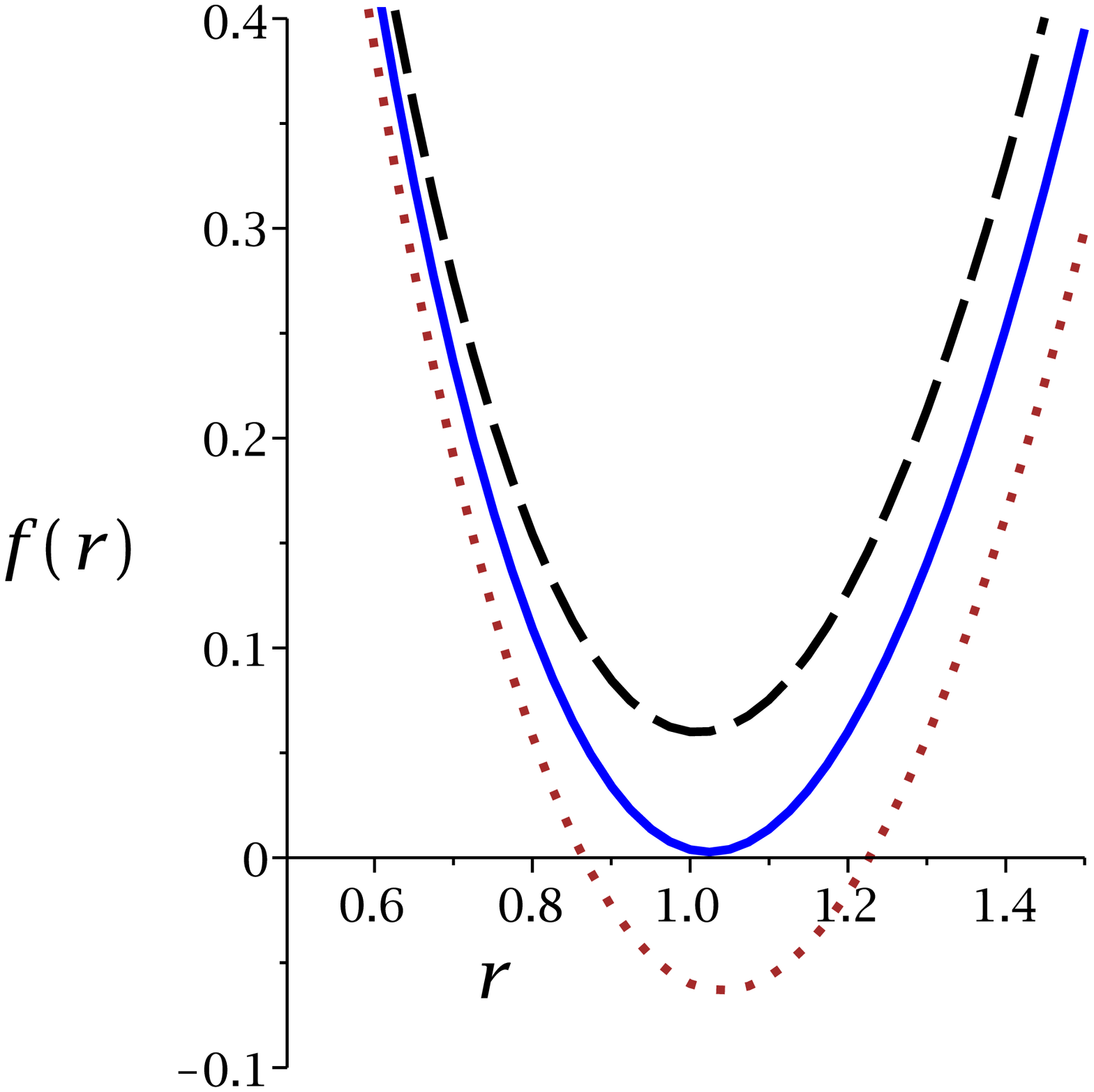} & \epsfxsize=5cm %
\epsffile{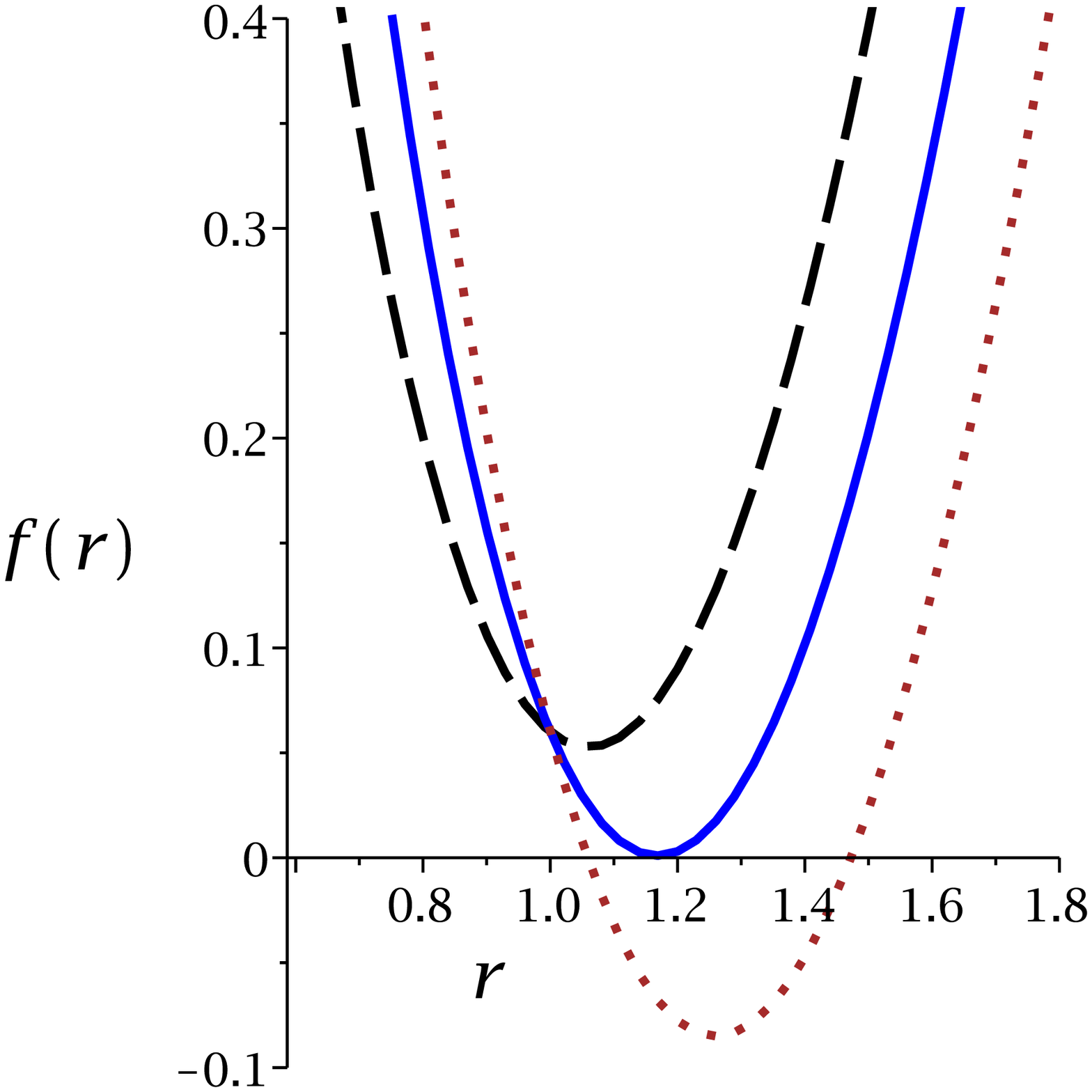} & \epsfxsize=5cm %
\epsffile{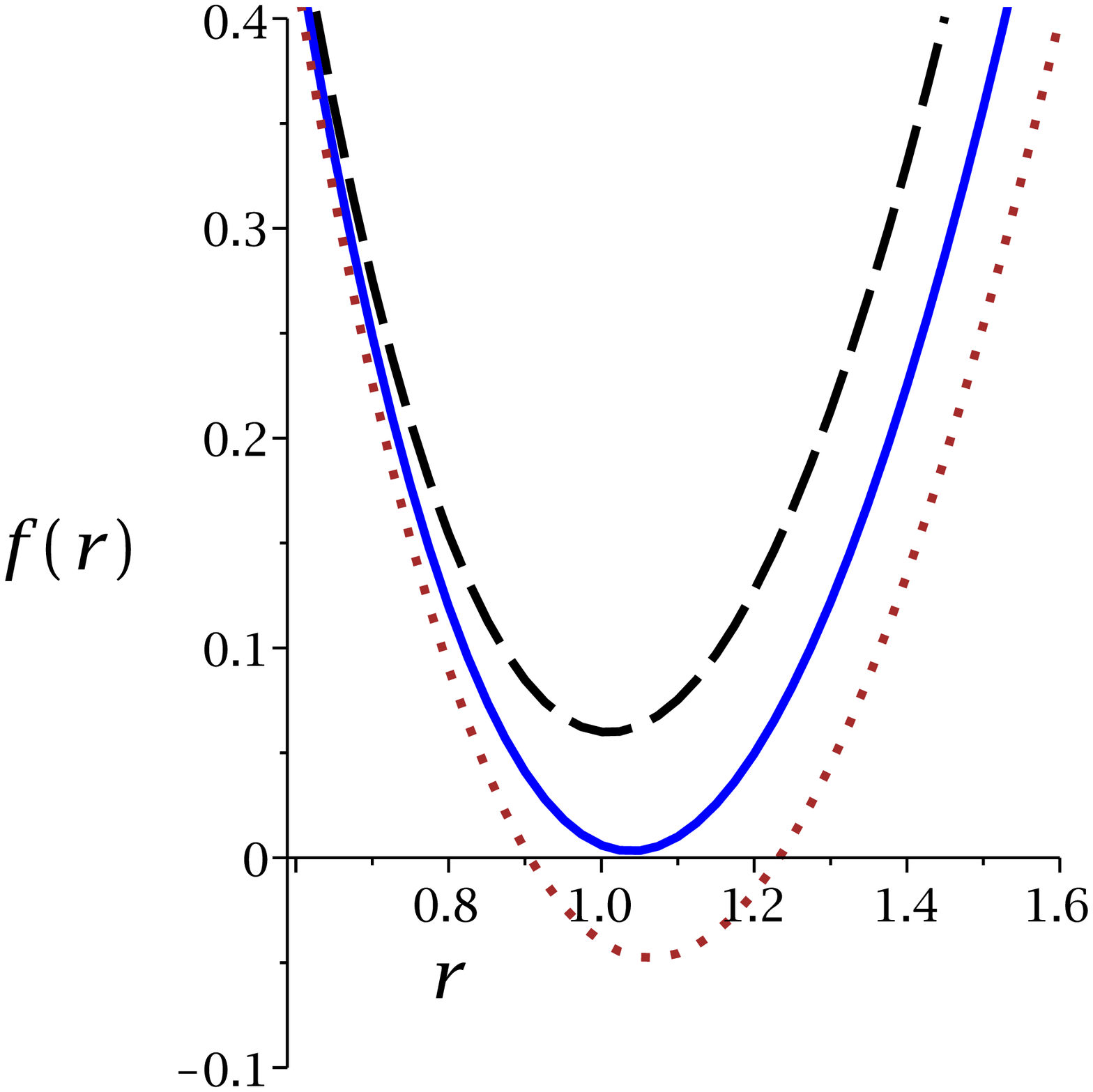}%
\end{array}
$%
\caption{\textbf{Eq. (\protect\ref{f(r)ENMax}):} $f(r)$ versus $r$ for $c=1$%
, $c_{1}=-1$, $m_{0}=0.9$ and $l=1$. \newline
Left panel: $\Lambda=-1$, $q=1$, $m=0.2$ (dashed line), $m=0.31$ (continuous
line) and $m=0.4$ (dotted line). \newline
Middle panel: $m=0.2$, $\Lambda=-1$, $q=1.05$ (dashed line), $q=1.16$
(continuous line) and $q=1.25$ (dotted line). \newline
Right panel: $q=1$, $m=0.2$, $\Lambda=-1$ (dashed line), $\Lambda=-0.95$
(continuous line) and $\Lambda=-0.9$ (dotted line).}
\label{Figfr2}
\end{figure}

In order to study the effects of free parameters on the metric function, we
can plot various diagrams (see Figs. \ref{Figfr1} and \ref{Figfr2}). These
figures show that by considering specific values for different parameters,
the metric function has different behaviors. The horizon properties of the
charged massive BTZ black hole may be like Reissner-Nordstr\"{o}m black
holes. In other words, this black hole may have two horizons, one extreme
horizon and without horizon (naked singularity) (see Figs. \ref{Figfr1} and %
\ref{Figfr2} for more details).

%%%%%%%%%%%%%%%%%%%%%%%%%%%%%%%%%%%%%%%%%%%%%%%%%%%%%%%%%%%%%%%

%%%%%%%%%%%%%%%%%%%%%%%%%%%%%%%%%%%%%%%%%%%%%%%%%%%%%%%%%%%%%%%
%%%%%%%%%%%%%%%%%%%%%%%%%%%%%%%%%%%%%%%%%%%%%%%%%%%%%%%%%%%%%%%

%%%%%%%%%%%%%%%%%%%%%%%%%%%%%%%%%%%%%%%%%%%%%%%%%%%%%%%%%%%%%%%

\subsection{Thermodynamics}

Now, we intend to calculate the conserved and thermodynamic quantities of
the solutions and examine the validity of the first law of thermodynamics.

Using the definition of Hawking temperature with relation to the surface
gravity on the outer horizon $r_{+}$, one can find
\begin{equation}
T=-\frac{\Lambda r_{+}}{2\pi }-\frac{q^{2}}{2\pi r_{+}}+\frac{m^{2}cc_{1}}{%
4\pi }.  \label{TotalTT}
\end{equation}

Using the Gauss's law, the electric charge, $Q$, can be found by calculating
the flux of the electric field at infinity, yielding
\begin{equation}
Q=\frac{q}{2}.  \label{TotalQ}
\end{equation}

Since we are working in Einstein gravity, the entropy of the black holes can
be obtained by employing the area law. It is a matter of calculation to show
that entropy has the following form \cite%
{hawking1,hawking2,hawking3,Hunter1,Hunter2,Hunter3}
\begin{equation}
S=\frac{\pi }{2}r_{+}.  \label{TotalS}
\end{equation}

By using the Hamiltonian approach and/or the counterterm method, one can
find the total mass of the solutions as
\begin{equation}
M=\frac{m_{0}}{8},  \label{TotalM}
\end{equation}
in which by evaluating metric function on the horizon ($f\left(
r=r_{+}\right) =0$), we obtain
\begin{equation}
M=-\Lambda r_{+}^{2}+m^{2}r_{+}cc_{1}-2q^{2}\ln \left( \frac{r_{+}}{l}%
\right).  \notag
\end{equation}

The electric potential, $U$, is defined through the gauge potential in the
following form
\begin{equation}
U_{Max}=A_{\mu }\chi ^{\mu }\left\vert _{r\rightarrow reference }\right.
-A_{\mu }\chi ^{\mu }\left\vert _{r\rightarrow r_{+}}\right. =-q\ln \left(
\frac{r_{+}}{l}\right)  \label{TotalU}
\end{equation}

Having conserved and thermodynamic quantities at hand, we are in a position
to check the validity of the first law of thermodynamics. It is easy to show
that by using thermodynamic quantities such as electric charge (\ref{TotalQ}%
), entropy (\ref{TotalS}) and mass (\ref{TotalM}), with the first law of
black hole thermodynamics
\begin{equation}
dM=TdS+UdQ,
\end{equation}
one can define the intensive parameters conjugate to $S$ and $Q$. These
quantities are the temperature and the electric potential
\begin{equation}
T=\left( \frac{\partial M}{\partial S}\right) _{Q}\ \ \ and\ \ \ \ \ \ \ \ \
U=\left( \frac{\partial M}{\partial Q}\right) _{S},  \label{TU}
\end{equation}
which are the same as those calculated for temperature (\ref{TotalTT}) and
electric potential (\ref{TotalU}). In other word, although massive term
modifies some of thermodynamic quantities, the first law of thermodynamics
is still valid.

\subsection{Thermal stability in the canonical ensemble}

Here, we study thermal stability criteria and the effects of different
parameters on such criteria. The stability conditions in canonical ensemble
are based on the sign of the heat capacity. This change of sign could happen
whether when heat capacity meets root(s) or divergency(ies). The root of
heat capacity (or temperature) indicates a bound point, which separates
physical solutions (positive temperature) from non-physical ones (negative
temperature). Whereas the roots of denominator of heat capacity (heat
capacity divergencies) represent phase transition points. The negativity of
heat capacity represents unstable solutions which may undergo a phase
transition and acquire stable state. In order to get a better picture and
enrich the results of our study, we investigate both temperature and heat
capacity, simultaneously.

The heat capacity is given by the following traditional relation
\begin{equation}
C_{Q}=\frac{T}{\left( \frac{\partial ^{2}M}{\partial S^{2}}\right) _{Q}}=%
\frac{T}{{\left( \frac{\partial T}{\partial S}\right) _{Q}}}.  \label{CQ}
\end{equation}

Considering Eqs. (\ref{TotalTT}) and (\ref{TotalS}), it is a matter of
calculation to show that
\begin{equation}
C_{Q}=\frac{2q^{2}+2\Lambda r_{+}^{2}-m^{2}cc_{1}r_{+}}{\left( \Lambda
r_{+}^{2}-q^{2}\right) }.  \label{CQMax}
\end{equation}

Obtained temperature (\ref{TotalTT}) consists three terms: cosmological
constant, electric charge and massive parameter. The electric charge term is
negative, therefore, its contribution is related to negativity of the
temperature. The contribution of the massive term depends on the sign of $%
c_{1}$. This term is not coupled with any order of horizon radius, which
results in it being constant. For adS spacetime, $\Lambda-$term is positive
and its contribution is toward positivity of temperature, the vise versa is
observed for dS spacetime. In conclusion, the dS black holes with negative $%
c_{1}$ has negative temperature every where, which results into absence of
physical black hole solutions for this case. Otherwise, the existence of
physical black holes (having positive temperature) is restricted to
satisfying a condition for massive term. For small values of horizon radius,
the dominant term is $q-$term while for large values of it, the dominant
term will be $\Lambda-$term. Therefore, for dS black holes, for small and
large values of the horizon radius, the temperature is negative and the
positive temperature may exist where massive term is dominant (for positive $%
c_{1}$). On the contrary, for adS black holes, for small values of horizon
radius temperature is negative while for sufficiently large values of
horizon radius, it is positive (regardless of massive term).

Now, we are in a position to study bound and phase transition points of
these black holes. Solving denominator and numerator of the heat capacity
with respect to horizon radius leads to following solutions for phase
transition and bound points, respectively,
\begin{equation}
r_{c}=\pm \frac{q}{\sqrt{\Lambda }}
\end{equation}
\begin{equation}
r_{0}=\frac{m^{2}cc_{1}\pm \sqrt{m^{4}c^{2}c_{1}^{2}-16\Lambda q^{2}}}{4
\Lambda }
\end{equation}

Interestingly, only for dS black holes phase transition exists and this
phase transition is independent of massive term. In other words, the
existence of second order phase transition is only a function of electric
charge and cosmological constant. On the contrary, the bound point is
observed for both dS and adS solutions. This bound point is a function of
the massive parameter, $\Lambda $ and electric charge. It is worthwhile to
mention that for the absence of massive gravity ($m=0$), only for adS
spacetime, bound point exists (no real bound point for dS black holes with $%
m=0$). This is the contribution of the massive gravity into thermodynamical
structure of the charged massive BTZ black holes. For adS black holes, only
one bound point is observed, whereas, for dS black holes, two bound points
could be obtained. Then again, we point out that existence of two bound
points for dS black holes is due to contribution of the massive gravity. It
is worthwhile to mention that in obtained heat capacity, massive term is
coupled with horizon radius. To finalize the study in this section, we
present various diagrams for temperature and heat capacity in case of
variation of different parameters for adS and dS spacetimes (see Figs. \ref%
{Fig1} - \ref{Fig2dS}).

%%%%%%%%%%%%%%%%%%%%%%%%%%%%%%%%%%%%%%%%%%%%%%%%%%%%%%%%%%%%%%%
\begin{figure}[tbp]
$%
\begin{array}{ccc}
\epsfxsize=5cm \epsffile{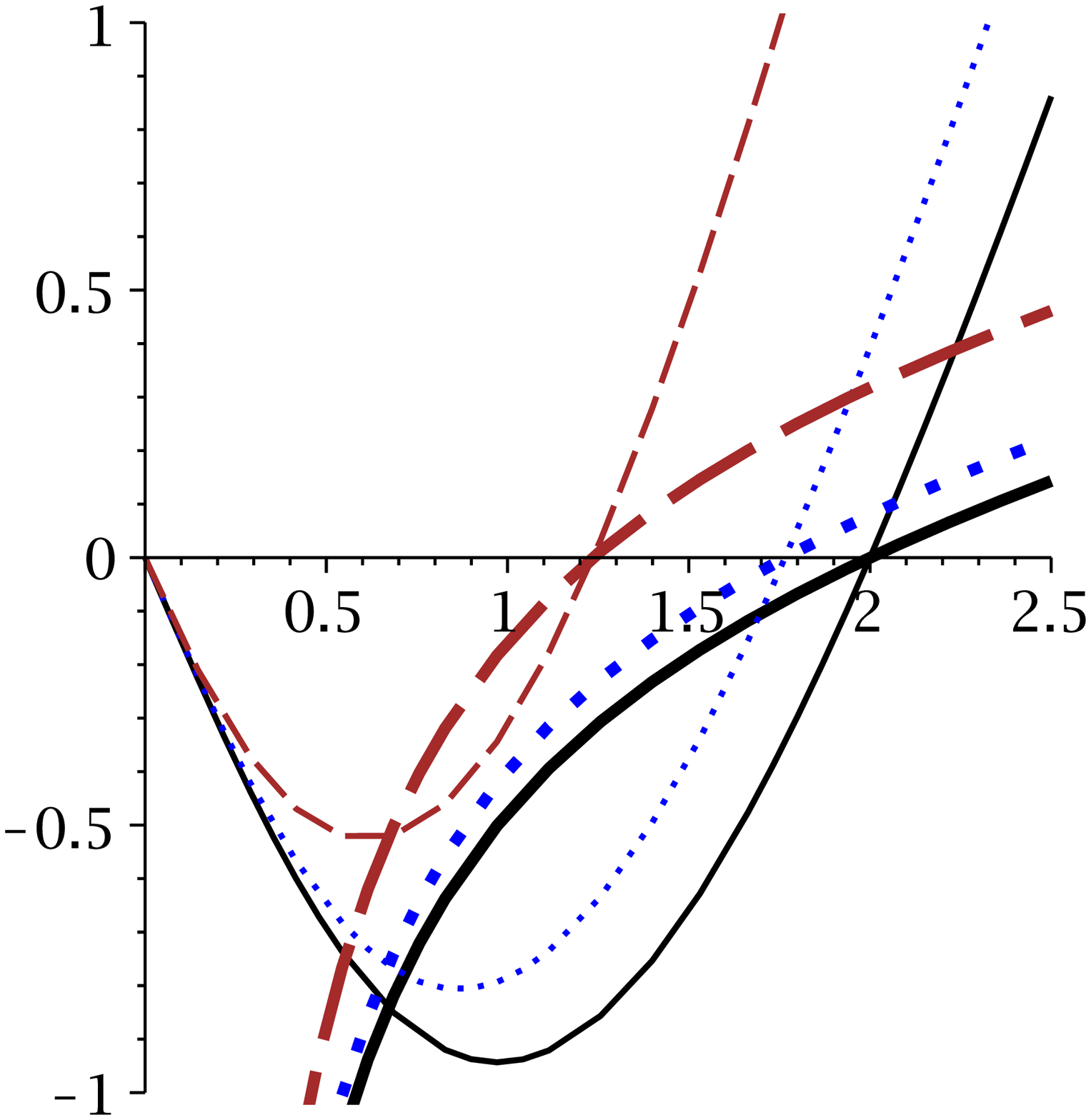} & \epsfxsize=5cm %
\epsffile{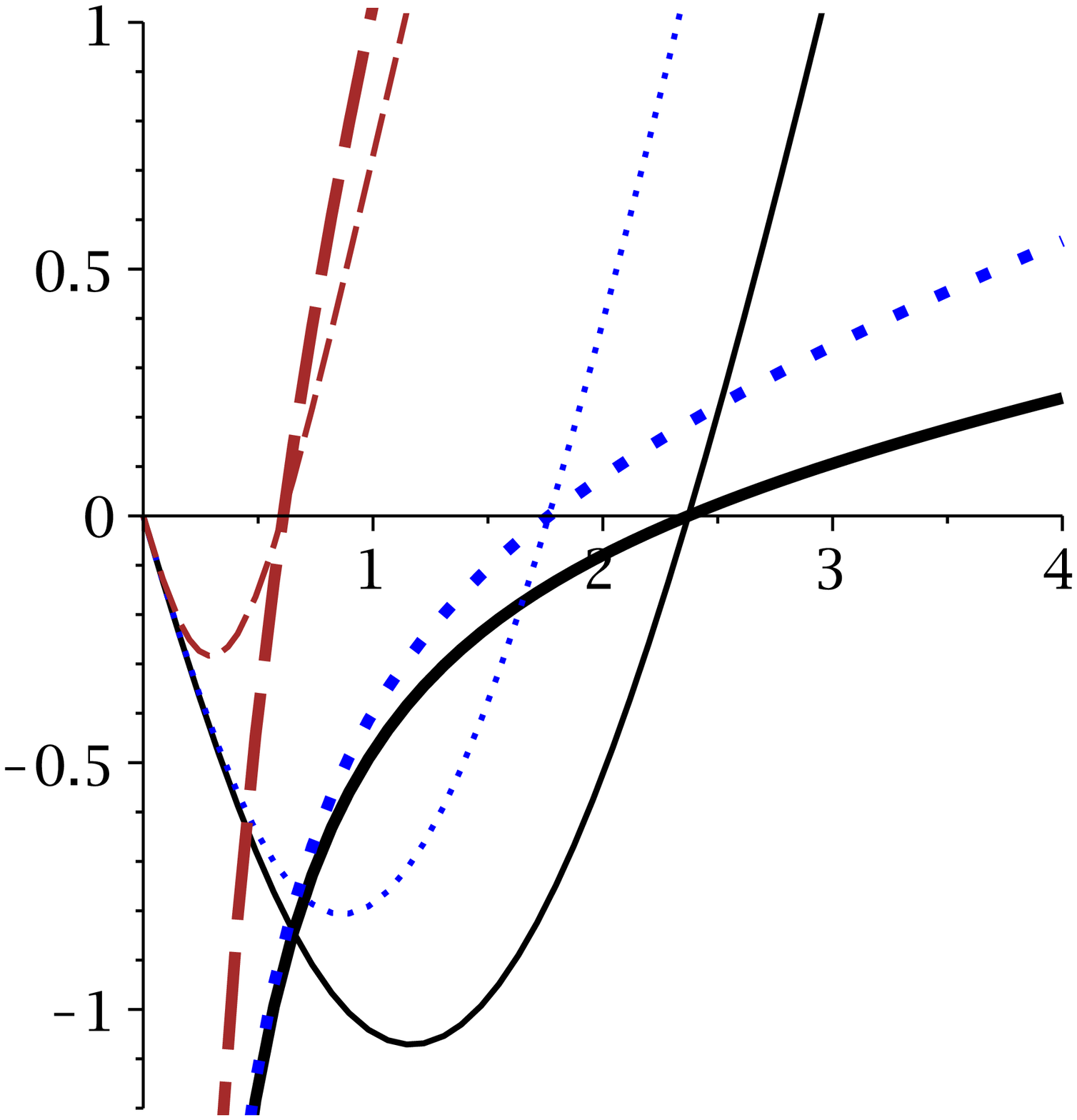} & \epsfxsize=5cm %
\epsffile{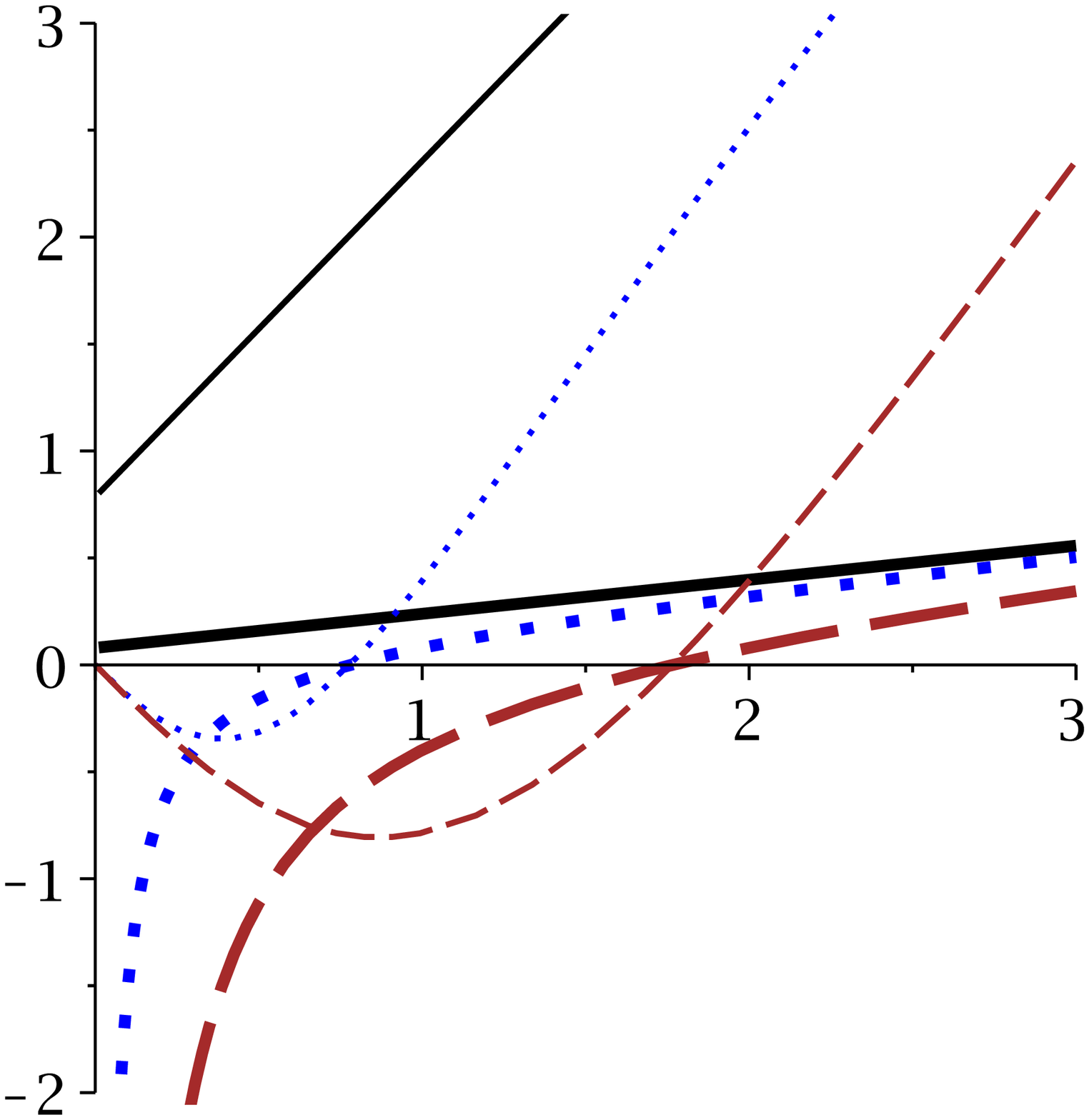}%
\end{array}
$%
\caption{$C_{Q}$ and $T$ (bold lines) versus $r_{+}$ for $c=c_{1}=1$.
\newline
Left panel: $q=2$, $\Lambda=-1$, $m=0$ (continuous line), $m=1$ (dotted
line) and $m=2$ (dashed line). \newline
Middle panel: $q=2$, $m=1$, $\Lambda=-0.5$ (continuous line), $\Lambda=-1$
(dotted line) and $\Lambda=-10$ (dashed line). \newline
Right panel: $\Lambda=-1$, $m=1$, $q=0$ (continuous line), $q=1$ (dotted
line) and $q=2$ (dashed line).}
\label{Fig1}
\end{figure}

%%%%%%%%%%%%%%%%%%%%%%%%%%%%%%%%%%%%%%%%%%%%%%%%%%%%%%%%%%%%%%%

%%%%%%%%%%%%%%%%%%%%%%%%%%%%%%%%%%%%%%%%%%%%%%%%%%%%%%%%%%%%%%%
\begin{figure}[tbp]
$%
\begin{array}{ccc}
\epsfxsize=5cm \epsffile{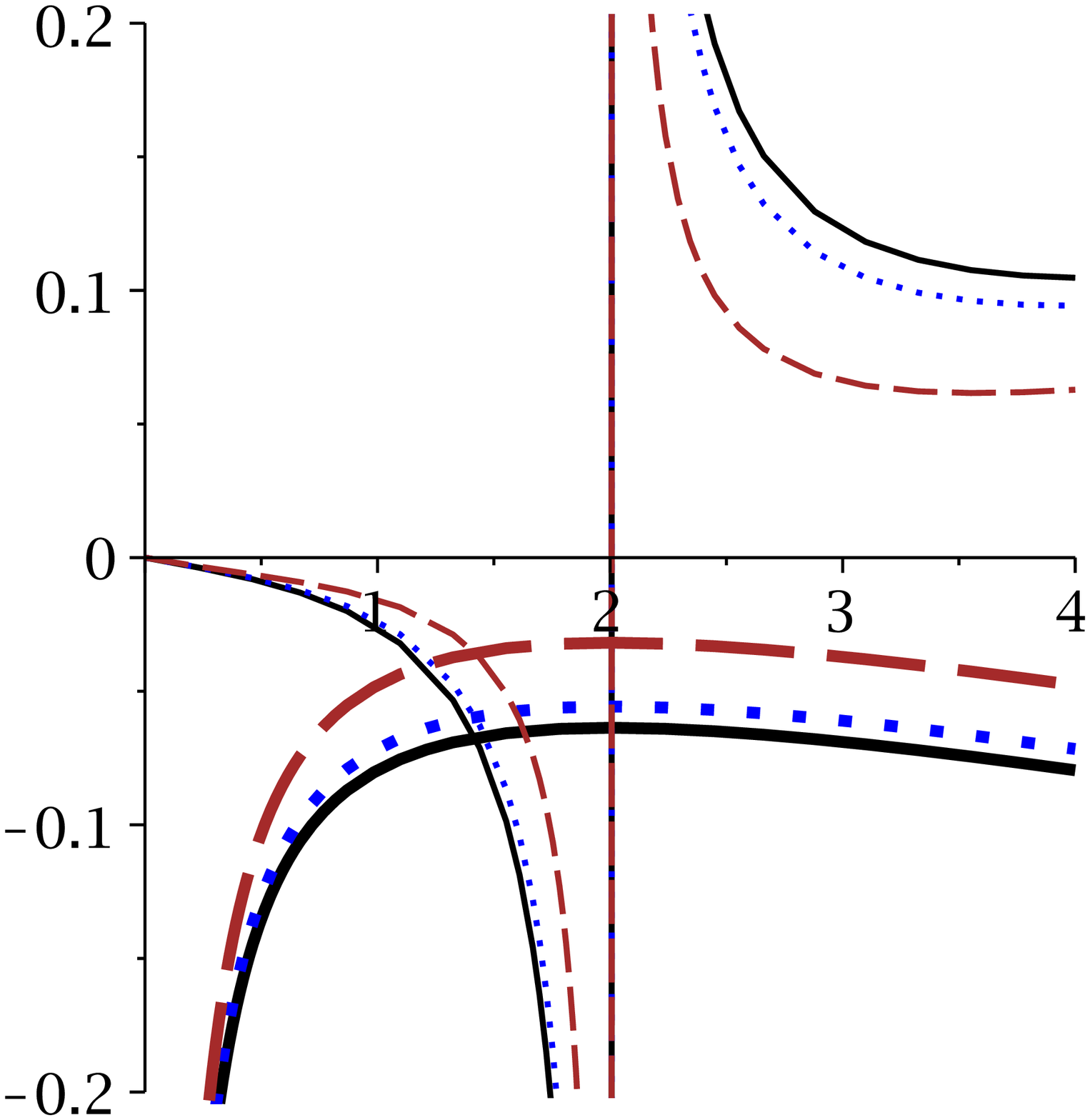} & \epsfxsize=5cm %
\epsffile{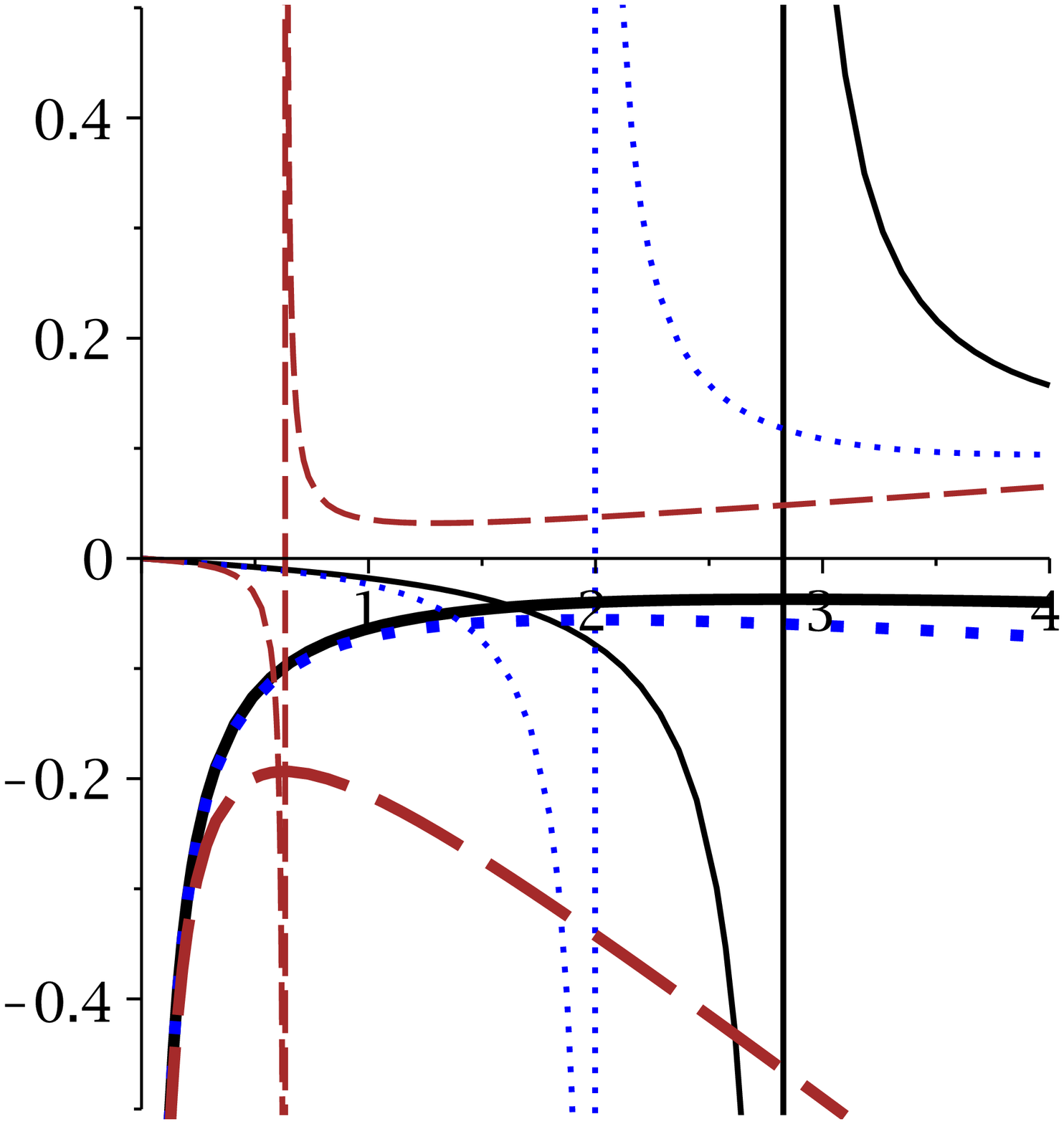} & \epsfxsize=5cm %
\epsffile{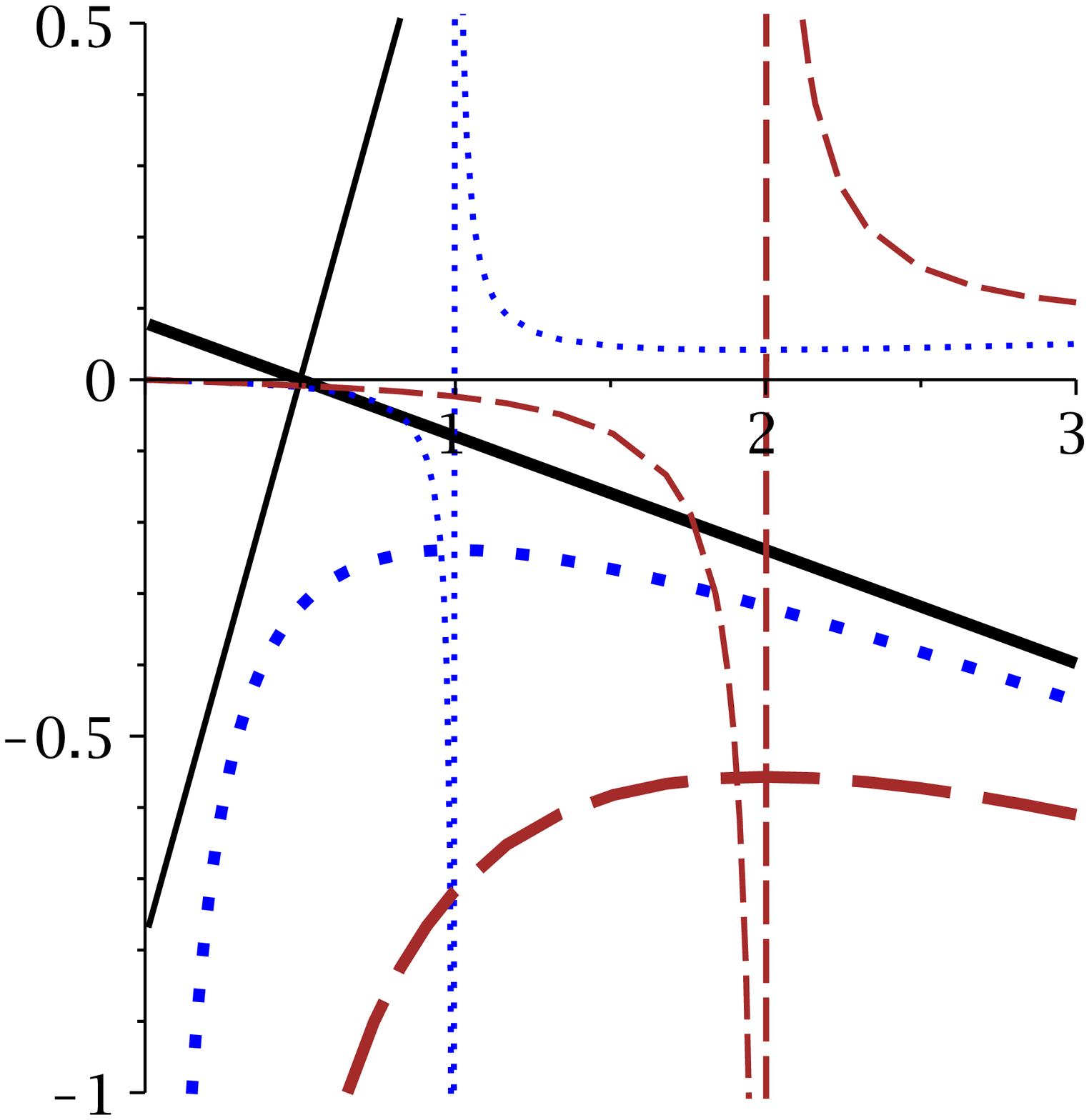}%
\end{array}
$%
\caption{For different scales: $C_{Q}$ and $T$ (bold lines) versus $r_{+}$
for $c=c_{1}=1$. \newline
Left panel: $q=2$, $\Lambda=1$, $m=0$ (continuous line), $m=1$ (dotted line)
and $m=2$ (dashed line). \newline
Middle panel: $q=2$, $m=1$, $\Lambda=0.5$ (continuous line), $\Lambda=1$
(dotted line) and $\Lambda=10$ (dashed line). \newline
Right panel: $\Lambda=1$, $m=1$, $q=0$ (continuous line), $q=1$ (dotted
line) and $q=2$ (dashed line).}
\label{Fig1dS}
\end{figure}

%%%%%%%%%%%%%%%%%%%%%%%%%%%%%%%%%%%%%%%%%%%%%%%%%%%%%%%%%%%%%%%
%%%%%%%%%%%%%%%%%%%%%%%%%%%%%%%%%%%%%%%%%%%%%%%%%%%%%%%%%%%%%%%
\begin{figure}[tbp]
$%
\begin{array}{ccc}
\epsfxsize=5cm \epsffile{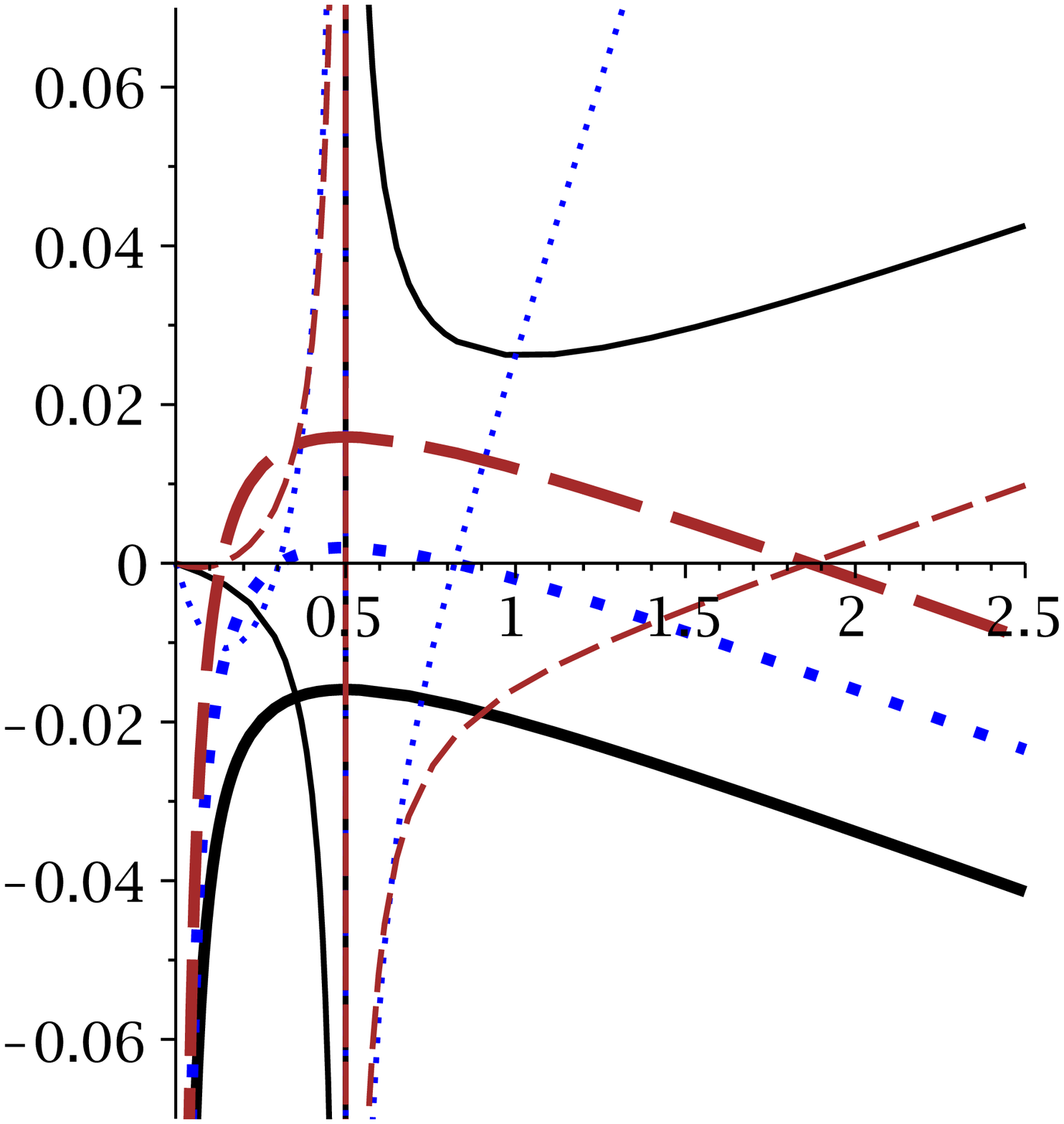} & \epsfxsize=5cm %
\epsffile{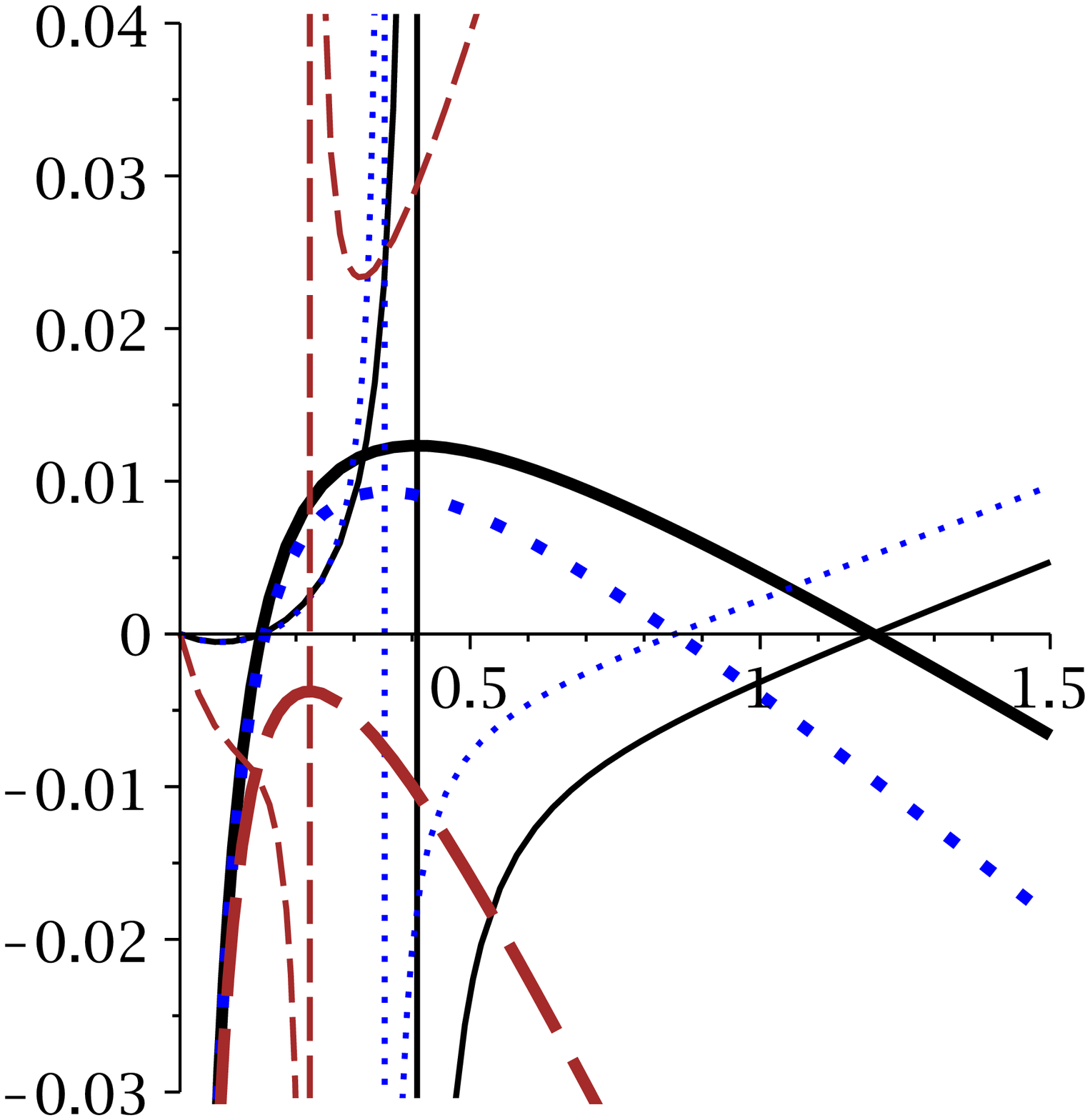} & \epsfxsize=5cm %
\epsffile{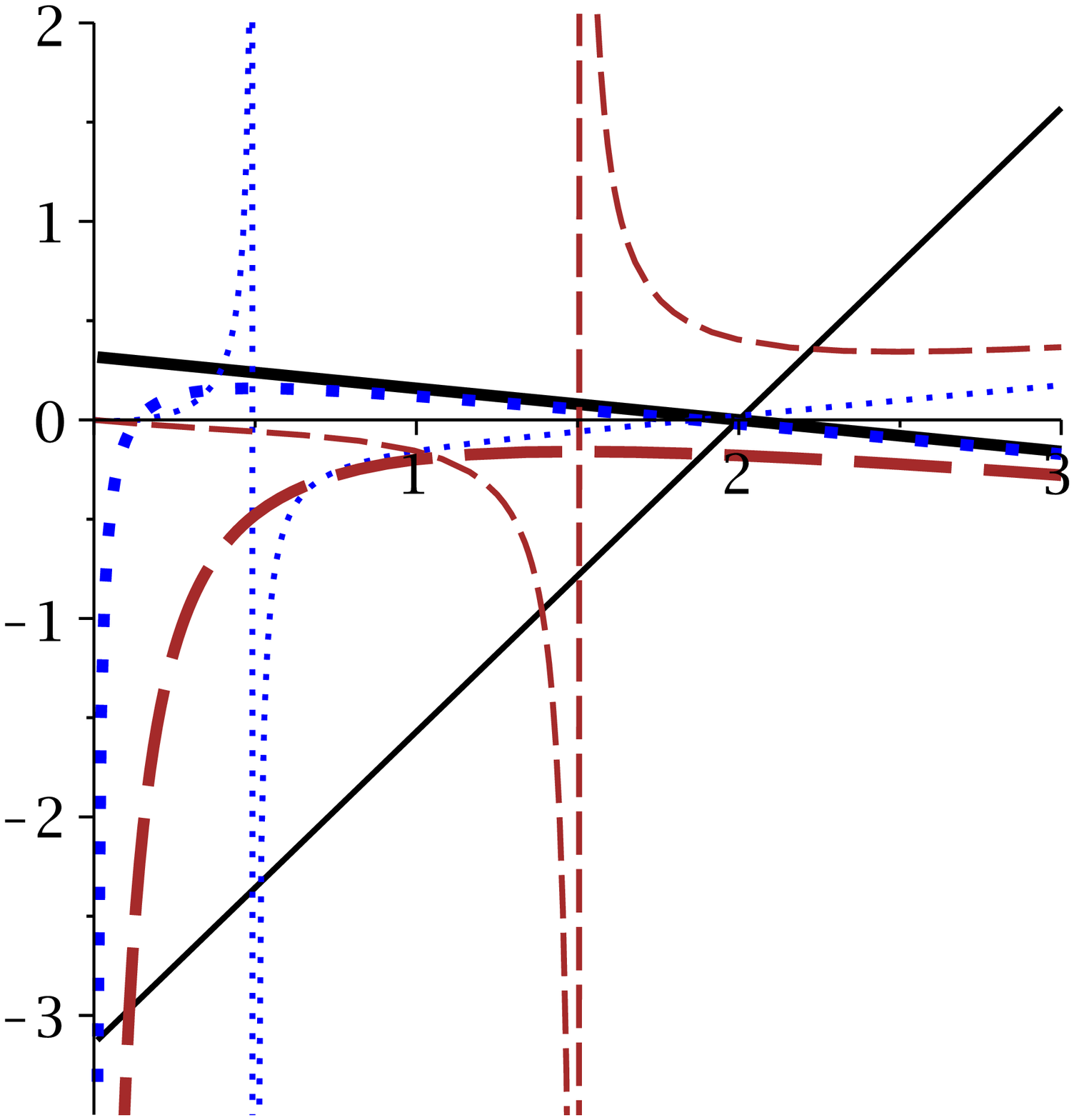}%
\end{array}
$%
\caption{For different scales: $C_{Q}$ and $T$ (bold lines) versus $r_{+}$
for $c=c_{1}=1$. \newline
Left panel: $q=0.5$, $\Lambda=1$, $m=0$ (continuous line), $m=1.4$ (dotted
line) and $m=2$ (dashed line). \newline
Middle panel: $q=0.5$, $m=2$, $\Lambda=1.5$ (continuous line), $\Lambda=2$
(dotted line) and $\Lambda=5$ (dashed line). \newline
Right panel: $\Lambda=1$, $m=2$, $q=0$ (continuous line), $q=0.5$ (dotted
line) and $q=1.5$ (dashed line).}
\label{Fig2dS}
\end{figure}

%%%%%%%%%%%%%%%%%%%%%%%%%%%%%%%%%%%%%%%%%%%%%%%%%%%%%%%%%%%%%%%

It is evident that for adS spacetime, bound point is an increasing function
of the electric charge (right panel of Fig. \ref{Fig1}) and a decreasing
function of massive parameter (left panel of Fig. \ref{Fig1}) and
cosmological constant (middle panel of Fig. \ref{Fig1}). For the absence of
electric charge, no bound point is observed and solutions are always
thermally stable.

For dS solutions, a second order phase transition exists (a divergency for
heat capacity is observed). For previous values of different parameters, we
only see non-physical solutions (see Fig. \ref{Fig1dS}). By suitable
choices, one can find physical solutions (see Fig. \ref{Fig2dS}). It is
evident that having positive temperature, hence, physical solutions is a
function of the massive parameter, cosmological constant and electric charge.

Increasing massive terms leads to formation of a physical region with two
bound points. Smaller bound point is a decreasing function of massive
parameter while the larger bound point is an increasing function of it (left
panel of Fig. \ref{Fig2dS}). On the contrary, increasing cosmological
constant and electric charge lead to absence of physical solutions. In other
words, smaller bound point is an increasing function of the cosmological
constat and electric charge whereas, the larger bound point is a decreasing
function of them (see middle and right panels of Fig. \ref{Fig2dS}). For
physical solutions, there is a phase transition of larger to smaller black
holes. In other words, even in physical region, only for specific range of
horizon radius thermally stable solutions exist. Once more, we emphasize
that existence of stable physical solutions for dS spacetime is only due to
contribution of the massive gravity.

\subsection{Neutral Massive Black Holes}

This subsection is devoted to the absence of the electric charge, since one
of the most interesting effects of the massive gravity and its interesting
phenomenology could be better observed in this case. Considering the case of
the absence of electric charge ($q=0$) and taking a look at the obtained
temperature and heat capacity, one can find
\begin{equation}
T=-\frac{\Lambda r_{+}}{2\pi }+\frac{m^{2}cc_{1}}{4\pi },  \label{Tq0}
\end{equation}%
\begin{equation}
C_{Q}=2-\frac{m^{2}cc_{1}}{\Lambda r_{+}}.
\end{equation}

The mentioned quantities confirm that both $T$ and $C_{Q}$ are positive for $%
\Lambda <0$ (adS solutions) while in order to have physical dS solutions ($%
\Lambda >0$) with positive temperature, one should set $r_{+}<\frac{%
m^{2}cc_{1}}{2\Lambda }$. This limitation on $r_{+}$ leads to obtain
negative $C_{Q}$, and therefore, we cannot obtain physically stable dS black
hole. Hereafter in this subsection, we consider adS solutions for more
discussions regarding vanishing event horizon.

Considering Eq. (\ref{Tq0}), one can find that for the limit $%
r_{+}\longrightarrow 0$, temperature of the black holes will not vanish. In
other words, for vanishing the horizon radius of the black holes, the $%
\Lambda-$term of temperature vanishes, whereas the massive term does not
vanish. Interestingly, the same behavior could be observed for heat capacity
as well. On the contrary, for $r_{+}=0$, total mass and entropy of black
holes will vanish.

Regarding black holes evaporation, we know that the mass and horizon radius
decrease during the evaporation process. One may regard the case $r_{+}=0$
as final state of black holes evaporation. The non-zero value of the
temperature for evaporation indicates that black holes leave a trace of
their existences even after complete evaporation. Therefore, all the
information, regarding the existence of black holes, is not actually lost in
evaporation. In fact, there is a contribution to temperature of the
background spacetime in place of the formation of black holes. The trace of
the existence of the black holes may be detected by this contribution to
temperature of the background spacetime. In other words, by measuring the
fluctuation of the temperature, one can determine formation and evaporation
of a black hole (a factor of $\frac{m^{2}cc_{1}}{4\pi }$ should be observed).

On the other hand, considering that thermodynamical temperature of the black
holes has geometrical interpretation as well. The non-zero temperature after
evaporation, indicates that information regarding the existence of the black
holes is stored and encoded in geometrical structure of the spacetime.
Meaning that, formation and existence of black holes altered spacetime
structure in a way that information regarding its existence is preserved.
This information (alteration in curvature and geometrical structure of the
spacetime) presents itself as a variation of the temperature in specific
place. Therefore, this variation of the temperature has no external source.
Its source is in the structure of spacetime itself.

\section{Einstein-Born-Infeld solutions in the context of massive gravity:}

\label{Einstein-Nonlinearly}

In this section, we obtain nonlinearly charged three dimensional solutions
through Born-Infeld theory in the context of massive gravity. Then, we study
their geometric and thermodynamic properties.

\subsection{Nonlinearly charged black hole solutions}

At first, we obtain Einstein-BI static black hole solutions in massive
gravity with (a)dS asymptotes. Using the ansatz metric (Eq. (\ref{f11})),
the $\mathcal{U}_{i}$'s are similar to Eq. (\ref{U}). Now, we consider BI
Lagrangian as matter field
\begin{equation}
L(\mathcal{F})=4\beta ^{2}\left( 1-\sqrt{1+\frac{\mathcal{F}}{2\beta ^{2}}}%
\right),  \label{LagBI}
\end{equation}
where $\beta$ is nonlinearity parameter. Considering the previous relation
of gauge potential ansatz (Eq. (\ref{gauge potential})), and using Eqs. (\ref%
{metric}), (\ref{Maxwell equation}) and (\ref{LagBI}), one finds
\begin{equation}
rh^{\prime \prime }(r)+h^{\prime }(r)\left[ 1-\left( \frac{h^{\prime }(r)}{%
\beta }\right) ^{2}\right] =0,  \label{eqh2}
\end{equation}
with the following solution
\begin{equation}
h(r)=q\ln \left( \frac{r}{2l}\left( 1+\Gamma \right) \right) ,
\label{h(r)II}
\end{equation}%
where $\Gamma =\sqrt{1+\frac{q^{2}}{r^{2}\beta ^{2}}}$. Also, the nonzero
components of electromagnetic field tensor are $F_{tr}=-F_{rt}=\frac{q}{%
r\Gamma }$.

Now, we use the introduced metric (Eq. \ref{metric}) and field equation of
Eq. (\ref{Field equation}) to obtain metric function $f(r)$. So,
respectively, $tt$ (or $rr$) and $\varphi \varphi$ components are in the
following forms
\begin{eqnarray}
&&f^{\prime }(r)+2\Lambda r+4r\beta ^{2}\left( \Gamma -1\right)
-m^{2}cc_{1}=0,  \label{eqENBI1} \\
&&f^{\prime \prime }(r)+2\Lambda -4\beta ^{2}\left( 1-\Gamma ^{-1}\right) =0.
\label{eqENBI2}
\end{eqnarray}

By employing Eqs. (\ref{eqENBI1}) and (\ref{eqENBI2}), one finds
\begin{equation}
f\left( r\right) _{BI}=-\Lambda r^{2}-m_{0}+2\beta ^{2}r^{2}\left( 1-\Gamma
\right) +q^{2}\left[ 1-2\ln \left( \frac{r}{2l}\left( 1+\Gamma \right)
\right) \right] +m^{2}cc_{1}r.  \label{f(r)Mass}
\end{equation}

It is easy to show that the obtained metric function (\ref{f(r)Mass}),
satisfy all components of Eq. (\ref{Field equation}), simultaneously. In
addition, such metric function reduces to Eq. (\ref{f(r)ENMax}) for $\beta
\rightarrow \infty$ (or large values of $r$). On the other hand, in the
absence of massive parameter ($m=0$), the solution (\ref{f(r)Mass}) reduces
to obtained black hole solution in \cite{HendiJHEP}
\begin{equation}
f\left( r\right) =-\Lambda r^{2}-m_{0}+2\beta ^{2}r^{2}\left( 1-\Gamma
\right) +q^{2}\left[ 1-2\ln \left( \frac{r}{2l}\left( 1+\Gamma \right)
\right) \right] .
\end{equation}

Here, we investigate the geometrical structure of this solution. For this
purpose, we look for the essential singularity(ies). The Ricci and the
Kretschmann scalars are
\begin{eqnarray}
R&=&6\Lambda -\frac{2m^{2}cc_{1}}{r}+\frac{2q^{2}}{r^{2}}+\frac{q^{4}}{%
2\beta ^{2}r^{4}}+O\left( \frac{1}{r^{6}}\right) , \\
&&  \notag \\
R_{\alpha \beta \gamma \delta }R^{\alpha \beta \gamma \delta } &=&12\Lambda
^{2}-\frac{8\Lambda m^{2}cc_{1}}{r}+\frac{2\left(
m^{4}c^{2}c_{1}^{2}+4\Lambda q^{2}\right) }{r^{2}}-\frac{8q^{2}m^{2}cc_{1}}{%
r^{3}}+\frac{12q^{4}}{r^{4}} \\
&&+\frac{2\Lambda q^{4}}{\beta ^{2}r^{4}}+\frac{2q^{4}m^{2}cc_{1}}{\beta
^{2}r^{5}}+O\left( \frac{1}{r^{6}}\right) .
\end{eqnarray}

Therefore at $r\longrightarrow 0$ we have
\begin{eqnarray}
\lim_{r\longrightarrow 0}R &\longrightarrow &\infty , \\
\lim_{r\longrightarrow 0}R_{\alpha \beta \gamma \delta }R^{\alpha \beta
\gamma \delta } &\longrightarrow &\infty ,
\end{eqnarray}%
which confirm that there is a curvature singularity at $r=0$. Also, the
Ricci and Kretschmann scalars are $6\Lambda$ and $12\Lambda^{2}$ at $%
r\longrightarrow \infty $ and hence, like linear solutions, the asymptotical
behavior of this solution is (a)dS for $\Lambda >0$ ($\Lambda <0$ ).

%%%%%%%%%%%%%%%%%%%%%%%%%%%%%%%%%%%%%%%%%%%%%%%%%%%%%%%%%%%%%%%

%%%%%%%%%%%%%%%%%%%%%%%%%%%%%%%%%%%%%%%%%%%%%%%%%%%%%%%%%%%%%%%
\begin{figure}[tbp]
$%
\begin{array}{ccc}
\epsfxsize=5cm \epsffile{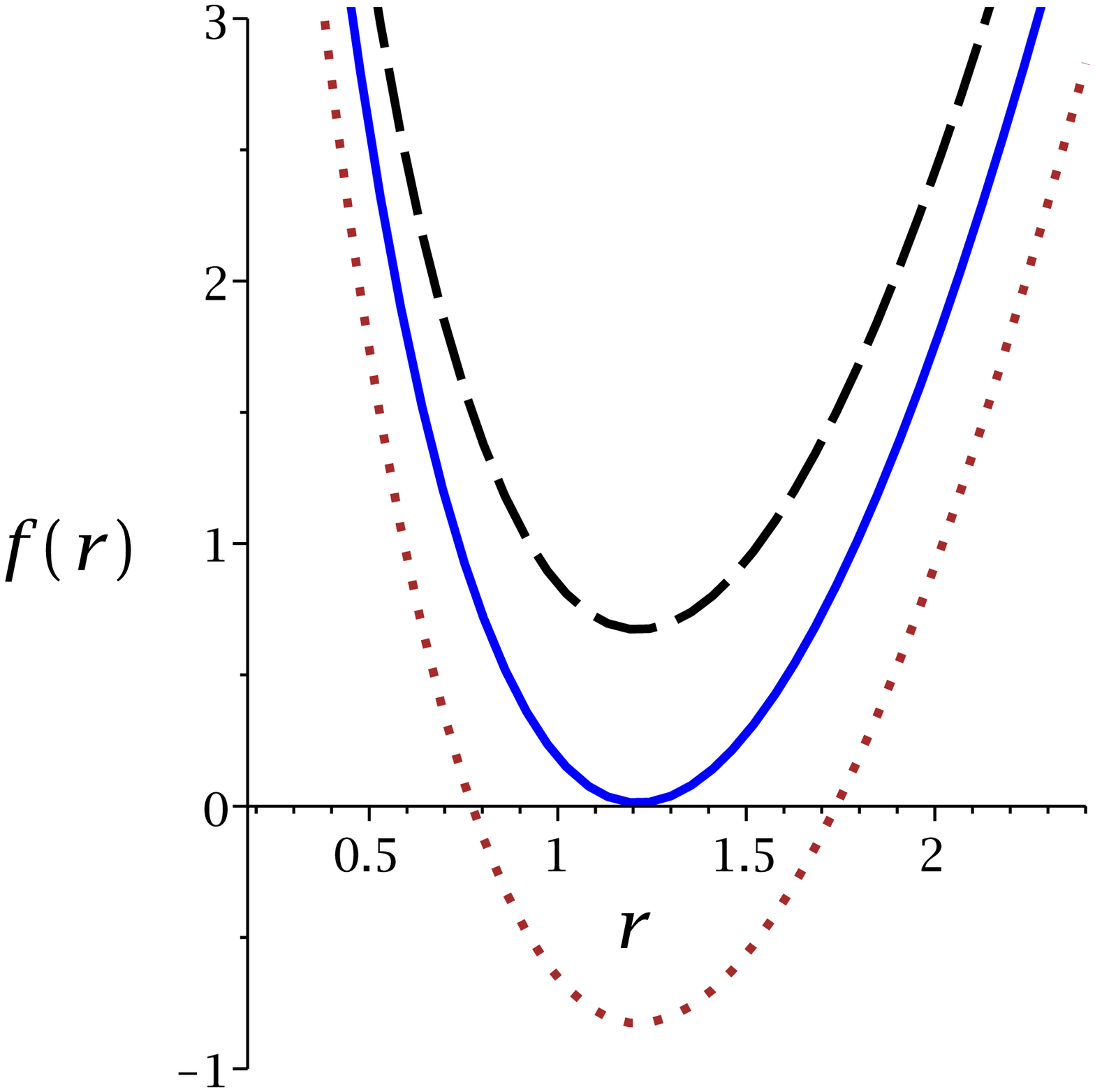} & \epsfxsize=5cm %
\epsffile{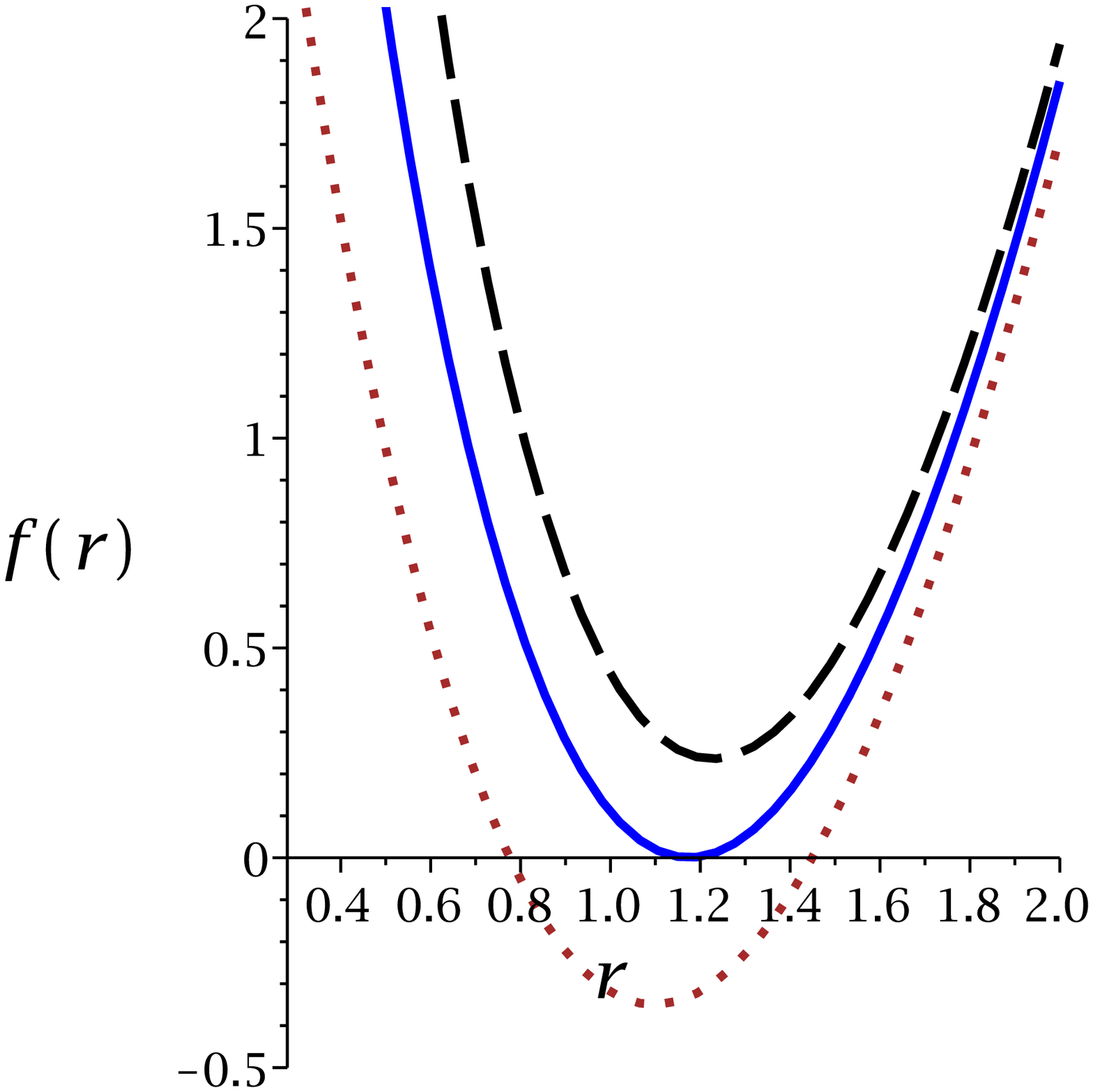} & \epsfxsize=5cm %
\epsffile{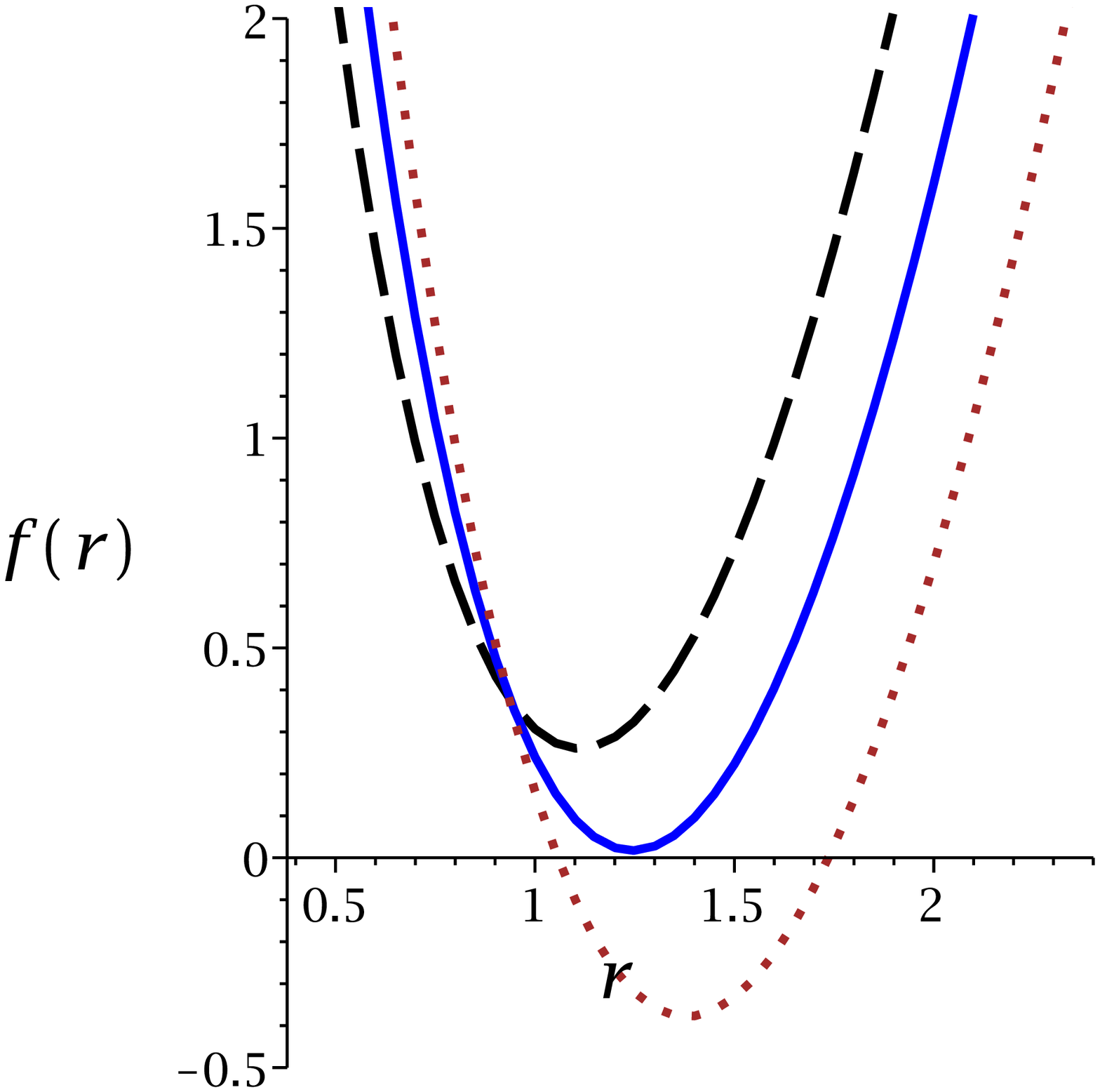}%
\end{array}
$%
\caption{$f(r)$ versus $r$ for $\Lambda=-1$, $c=1$, $c_{1}=1$, $m=2$ and $%
l=1 $. \newline
Left panel: $q=2$, $\protect\beta=5$, $m_{0}=4.00$ (dashed line), $%
m_{0}=4.66 $ (continuous line) and $m_{0}=5.50$ (dotted line). \newline
Middle panel: $q=2$, $m_{0}=4.5$, $\protect\beta=8.00$ (dashed line), $%
\protect\beta=3.05$ (continuous line) and $\protect\beta=2.00$ (dotted
line). \newline
Right panel: $\protect\beta=4$, $m=4.5$, $q=1.90$ (dashed line), $q=2.05$
(continuous line) and $q=2.20$ (dotted line).}
\label{Figfr3}
\end{figure}

%%%%%%%%%%%%%%%%%%%%%%%%%%%%%%%%%%%%%%%%%%%%%%%%%%%%%%%%%%%%%%%

By considering specific values for different parameters, the metric function
has different behaviors. The obtained black holes may behave like
Reissner-Nordstr\"{o}m black holes, and hence, the solution may be black
hole with two horizons, black hole with one extreme horizon and naked
singularity (see Fig. \ref{Figfr3} for more details). In addition, this
nonlinear solution has additional properties. Following the approach of Ref.
\cite{HendiJHEP}, one finds a critical value for the nonlinearity parameter (%
$\beta_{c}$), in which for $\beta<\beta_{c}$, the nonlinear solution behaves
like Schwarzschild black hole (singularity cover with a non-extreme horizon).

\subsection{Thermodynamics}

Here, we calculate the conserved and thermodynamic quantities of the
nonlinearly charged black hole solution and then we check the validity of
the first law of thermodynamics.

Using the definition of surface gravity on the outer horizon $r_{+}$
(Hawking temperature), one finds
\begin{equation}
T=-\frac{\Lambda r_{+}}{2\pi }-\frac{q^{2}}{\pi r_{+}\left( 1+\Gamma_{+}
\right) }+\frac{m^{2}cc_{1}}{4\pi },  \label{TBI}
\end{equation}%
where $\Gamma _{+}=\sqrt{1+\frac{q^{2}}{r_{+}^{2}\beta ^{2}}}$. Calculations
regarding the total electric charge, entropy and total mass show that the
charge, the entropy and the mass of this black hole are similar to Eqs. (\ref%
{TotalQ})--(\ref{TotalM}), respectively. Evaluating metric function on the
horizon ($f\left( r=r_{+}\right) =0$), one can obtain total mass as a
function of different parameters as
\begin{equation}
M=-\frac{\Lambda r_{+}^{2}}{8}-\frac{q^{2}\ln \left[ \frac{r_{+}}{2l}\left(
1+\Gamma _{+}\right) \right] }{4}+\frac{2r_{+}^{2}\beta ^{2}\left(
1-\Gamma_{+} \right) +q^{2}+m^{2}cc_{1}r_{+}}{8}.  \label{MBI}
\end{equation}

In addition, the electric potential, $U$, is in the following form
\begin{equation}
U=-q\ln \left[ \frac{r_{+}}{2l}\left( 1+\Gamma_{+} \right) \right] .
\label{UBI}
\end{equation}

Using obtained electric charge and potential with entropy, temperature and
total mass, it is straightforward to examine the validity of the first law
of black hole thermodynamics as
\begin{equation}
dM=TdS+UdQ.
\end{equation}

It is worthwhile to mention that although nonlinearity and massive gravity
modify some of thermodynamic quantities, the first law remains valid.

\subsection{Heat capacity and thermal stability}

In this subsection, we investigate the effects of BI generalization of
BTZ-massive black hole on thermodynamical stability through heat capacity.
Considering Eq. (\ref{CQ}) with entropy (\ref{TotalS}) and temperature (\ref%
{TBI}) of BI black hole, one can find
\begin{equation}
C_{Q}=\frac{\pi r_{+}^{2}\beta ^{2}\left[ \left( m^{2}cc_{1}-2\Lambda
r_{+}\right) \left( 1+\Gamma _{+}\right) -\frac{4q^{2}}{r_{+}}\right] \Gamma
_{+}}{4\left[ q^{2}\left( \Lambda -2\beta ^{2}\right) +\Lambda
r_{+}^{2}\beta ^{2}\left( 1+\Gamma _{+}\right) \right] }.  \label{CQBI}
\end{equation}

Solving numerator and denominator of the obtained heat capacity, one can
find following relations for bound and phase transition points, respectively
\begin{eqnarray}
r_{0} &=&\frac{m^{2}cc_{1}\left( \Lambda -2\beta ^{2}\right) \pm \beta \sqrt{%
\beta ^{2}m^{4}c^{2}c_{1}^{2}-4\Lambda q^{2}\left( \Lambda -4\beta
^{2}\right) }}{\Lambda \left( \Lambda -4\beta ^{2}\right) },  \label{r0BI} \\
&&  \notag \\
r_{c} &=&\pm \frac{\left( \Lambda -2\beta ^{2}\right) q}{\beta \sqrt{\Lambda
\left( 4\beta ^{2}-\Lambda \right) }}.  \label{rcBI}
\end{eqnarray}

\begin{figure}[tbp]
$%
\begin{array}{cc}
\epsfxsize=6.5cm \epsffile{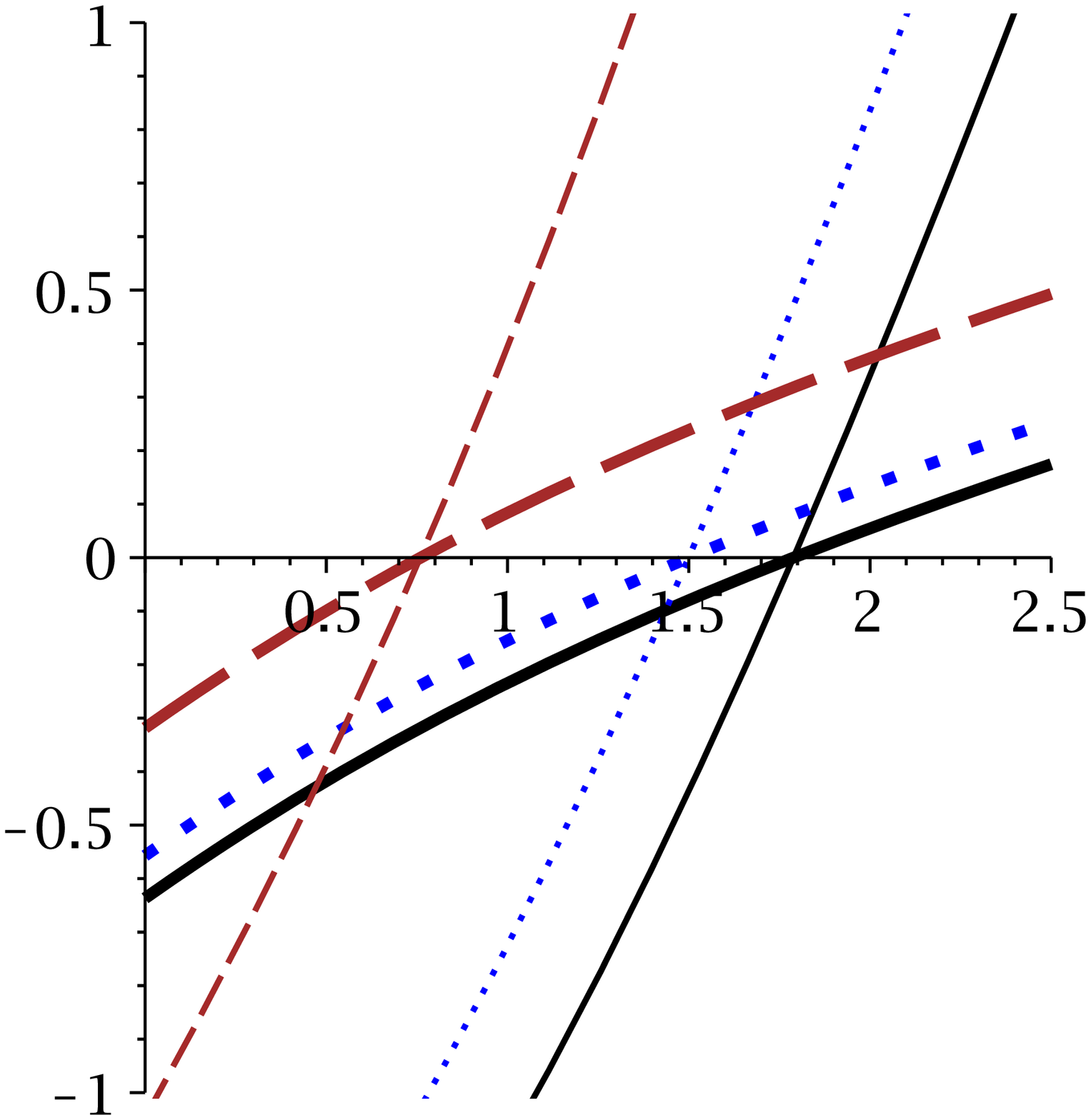} & \epsfxsize=6.5cm %
\epsffile{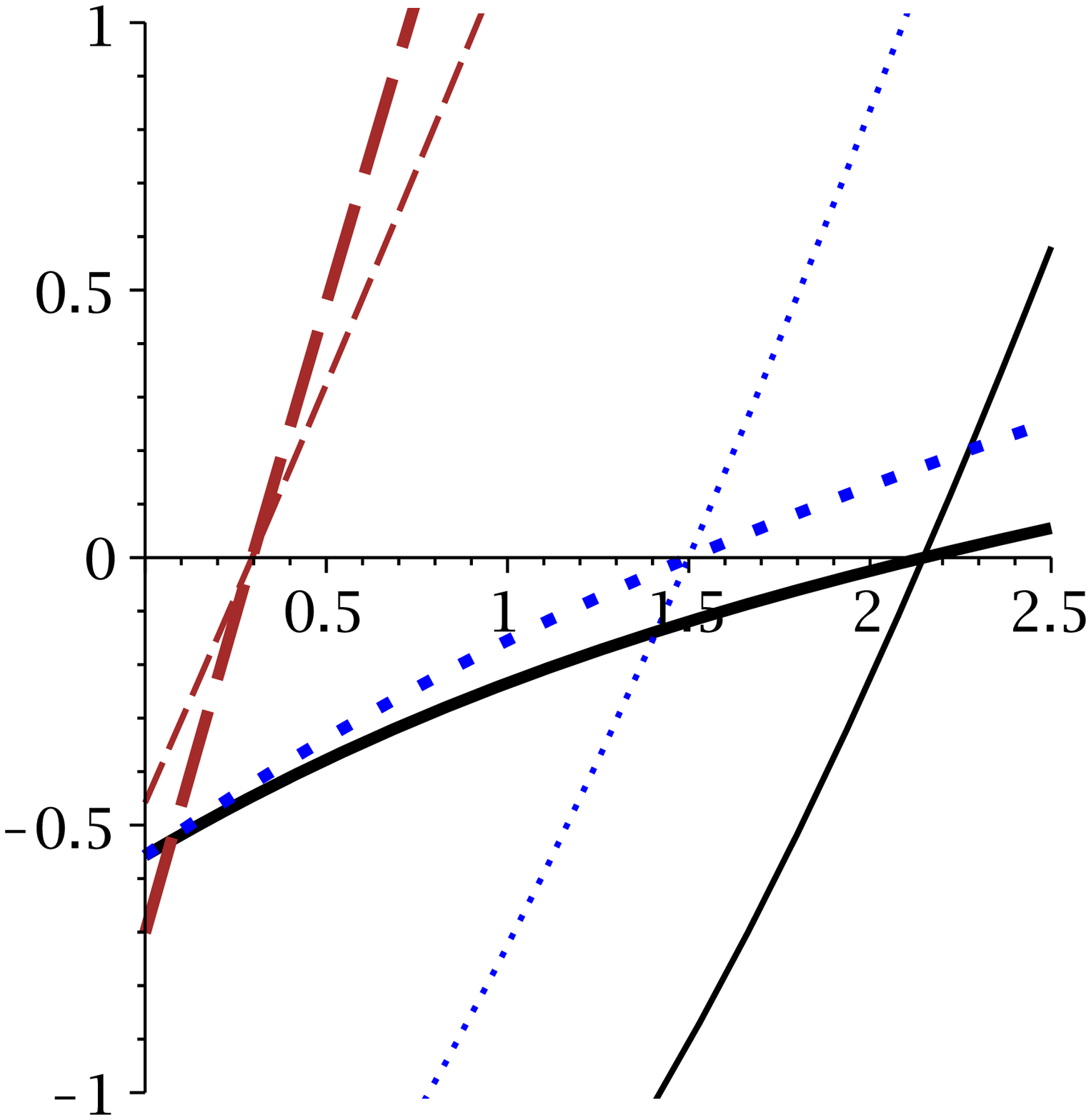}%
\end{array}
$%
\caption{$C_{Q}$ and $T$ (bold lines) versus $r_{+}$ for $c=c_{1}=1$, $q=2$
and $\protect\beta=1$. \newline
Left panel: $\Lambda=-1$, $m=0$ (continuous line), $m=1$ (dotted line) and $%
m=2$ (dashed line). \newline
Right panel: $m=1$, $\Lambda=-0.5$ (continuous line), $\Lambda=-1$ (dotted
line) and $\Lambda=-10$ (dashed line).}
\label{Fig2}
\end{figure}

\begin{figure}[tbp]
$%
\begin{array}{cc}
\epsfxsize=6.5cm \epsffile{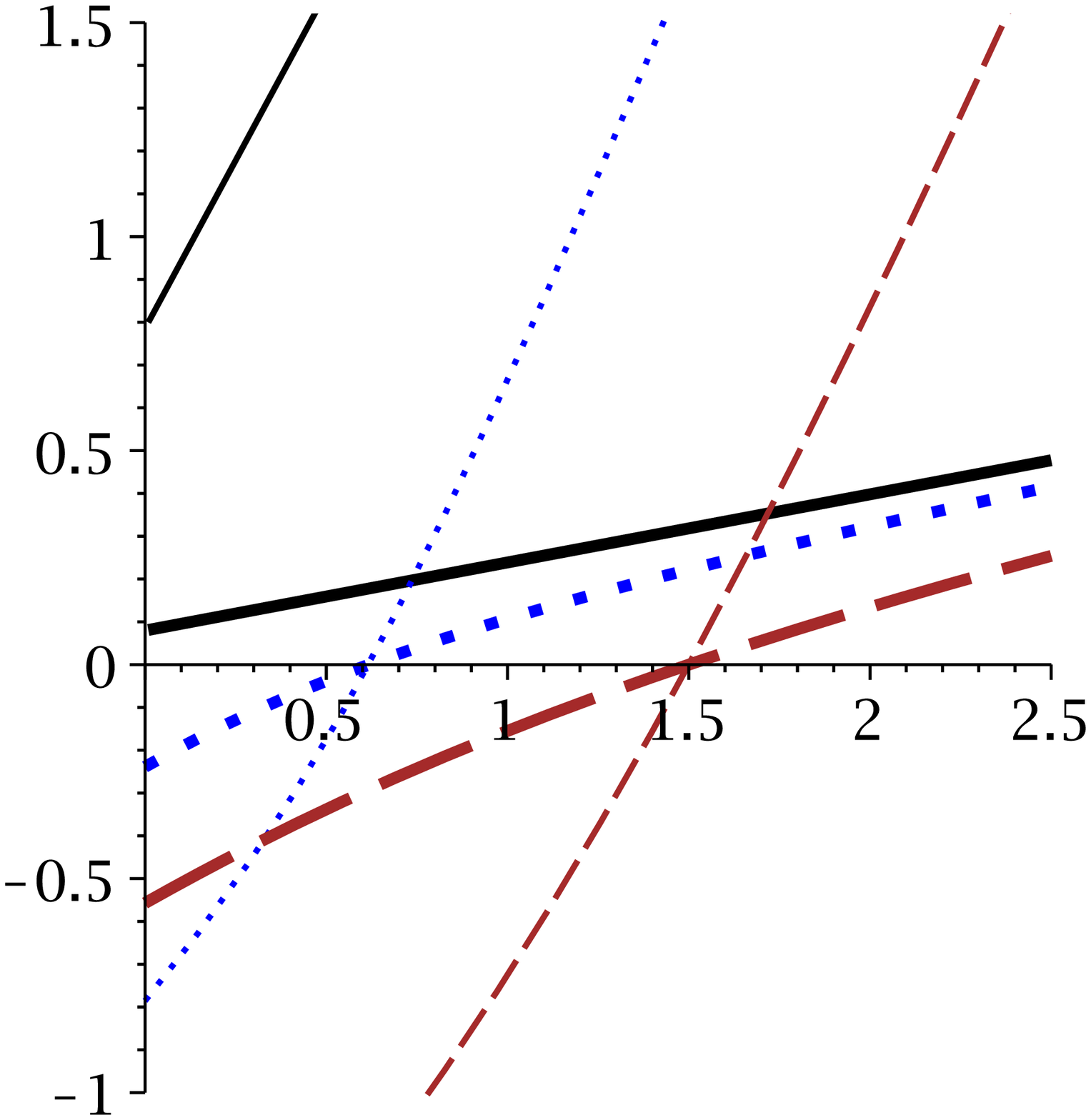} & \epsfxsize=6.5cm %
\epsffile{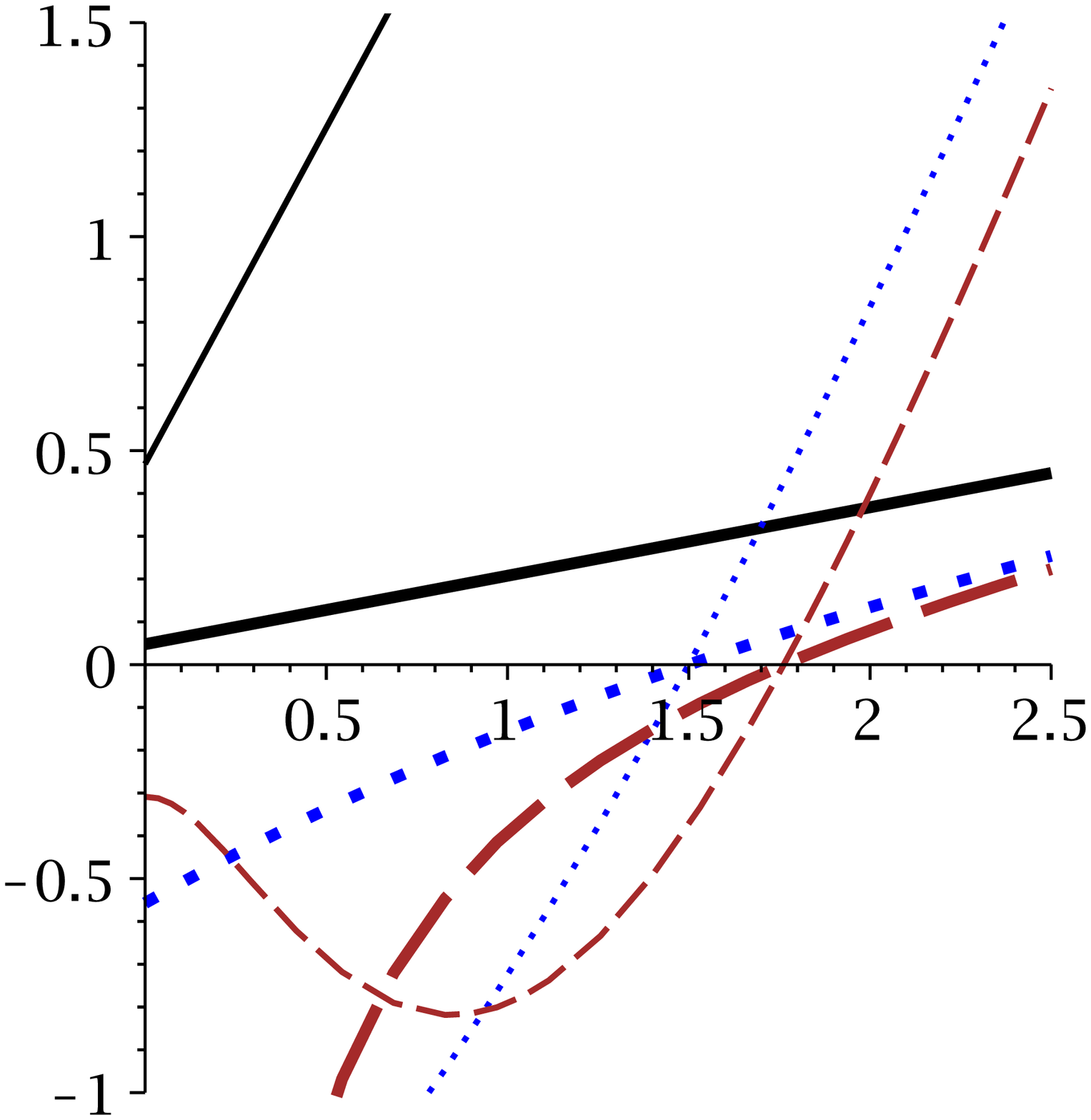}%
\end{array}
$%
\caption{$C_{Q}$ and $T$ (bold lines) versus $r_{+}$ for $c=c_{1}=1$, $%
\Lambda=-1$ and $m=1$. \newline
Left panel: $\protect\beta=1$, $q=0$ (continuous line), $q=1$ (dotted line)
and $q=2$ (dashed line). \newline
Right panel: $q=2$, $\protect\beta=0.05$ (continuous line), $\protect\beta=1$
(dotted line) and $\protect\beta=10$ (dashed line).}
\label{Fig3}
\end{figure}

\begin{figure}[tbp]
$%
\begin{array}{cc}
\epsfxsize=6.5cm \epsffile{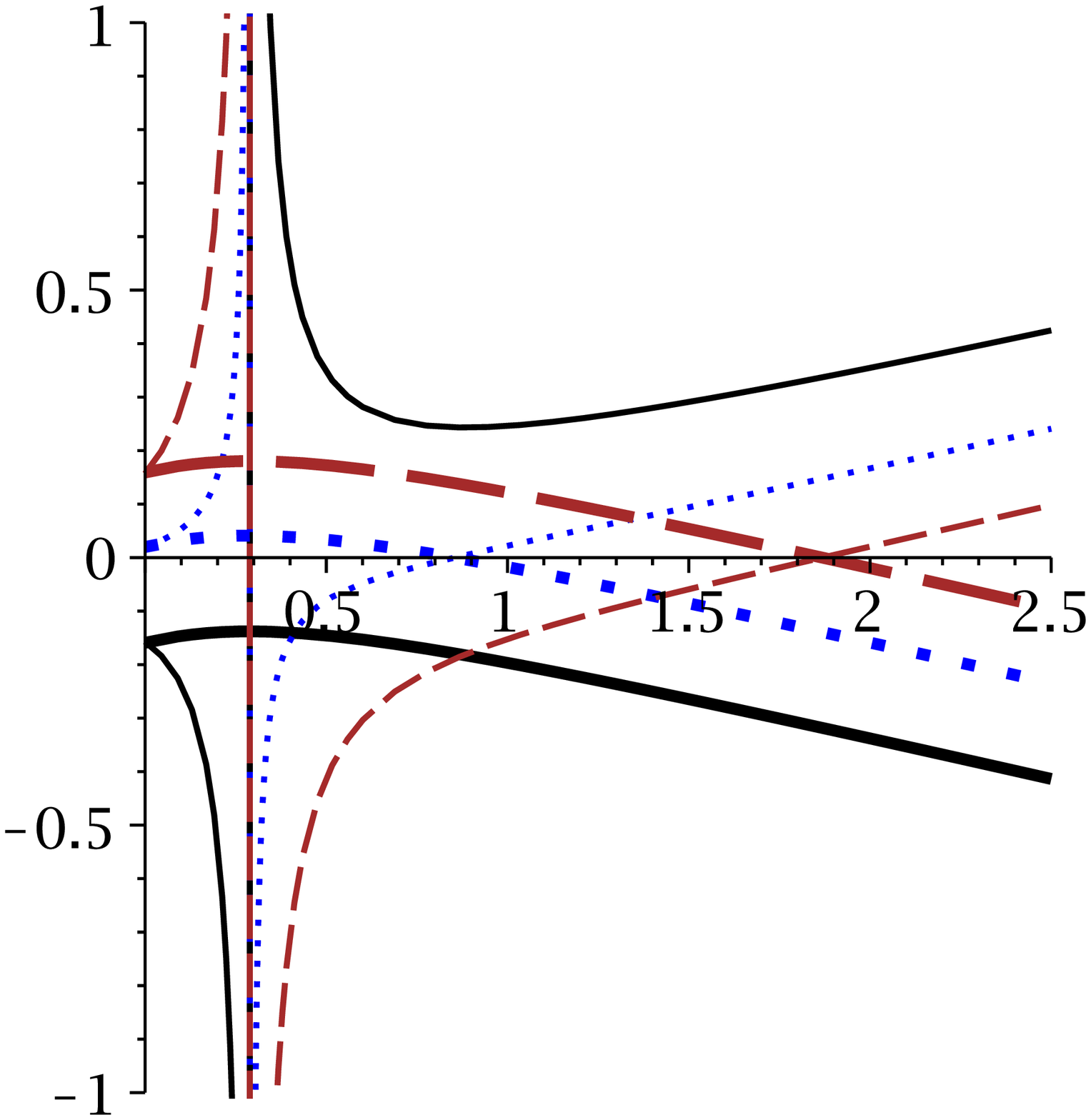} & \epsfxsize=6.5cm %
\epsffile{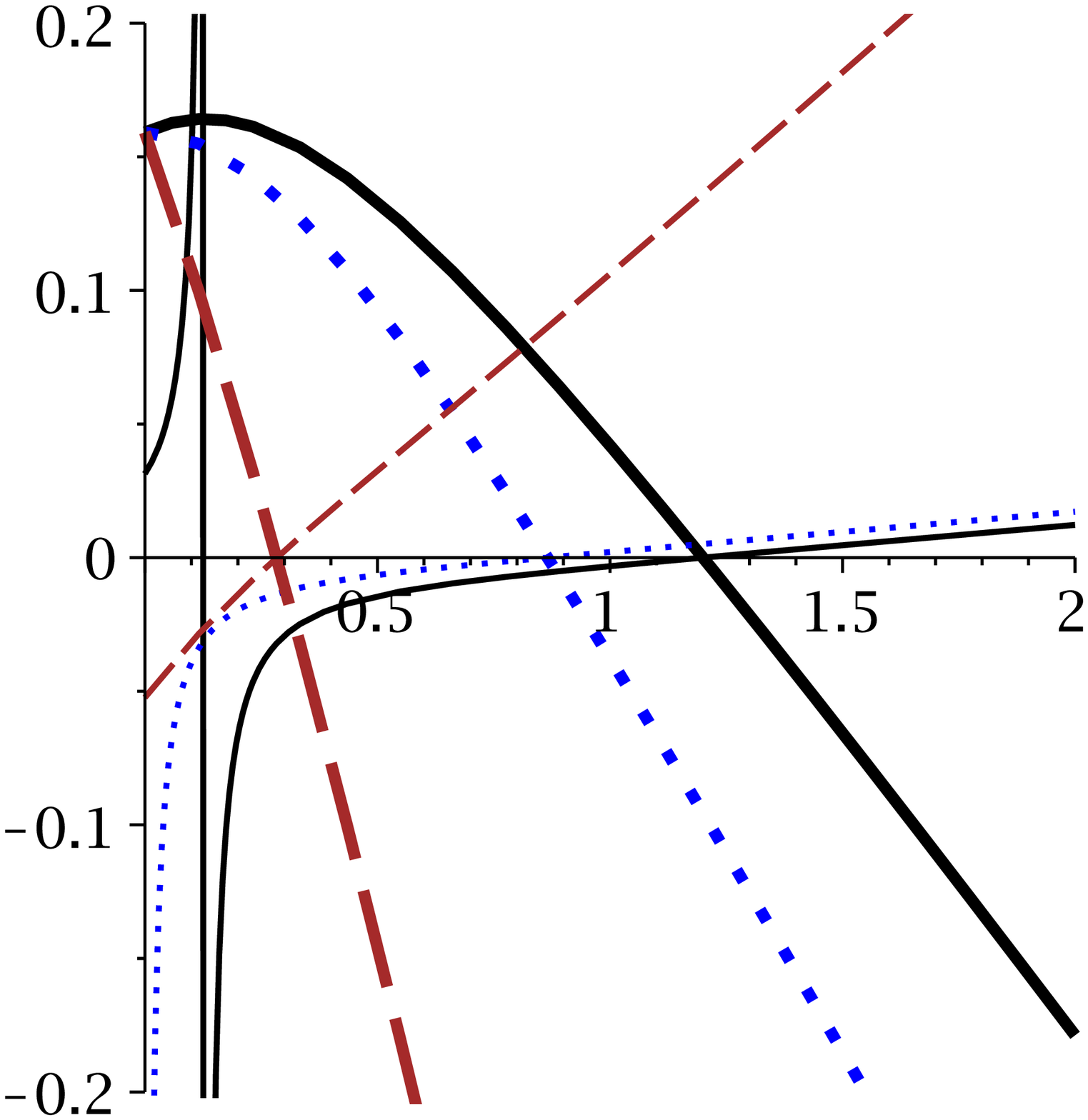}%
\end{array}
$%
\caption{For different scales: $C_{Q}$ and $T$ (bold lines) versus $r_{+}$
for $c=c_{1}=1$, $q=0.5$ and $\protect\beta=1$. \newline
Left panel: $\Lambda=1$, $m=0$ (continuous line), $m=1.5$ (dotted line) and $%
m=2$ (dashed line). \newline
Right panel: $m=2$, $\Lambda=1.5$ (continuous line), $\Lambda=2$ (dotted
line) and $\Lambda=5$ (dashed line).}
\label{Fig22dS}
\end{figure}

\begin{figure}[tbp]
$%
\begin{array}{cc}
\epsfxsize=6.5cm \epsffile{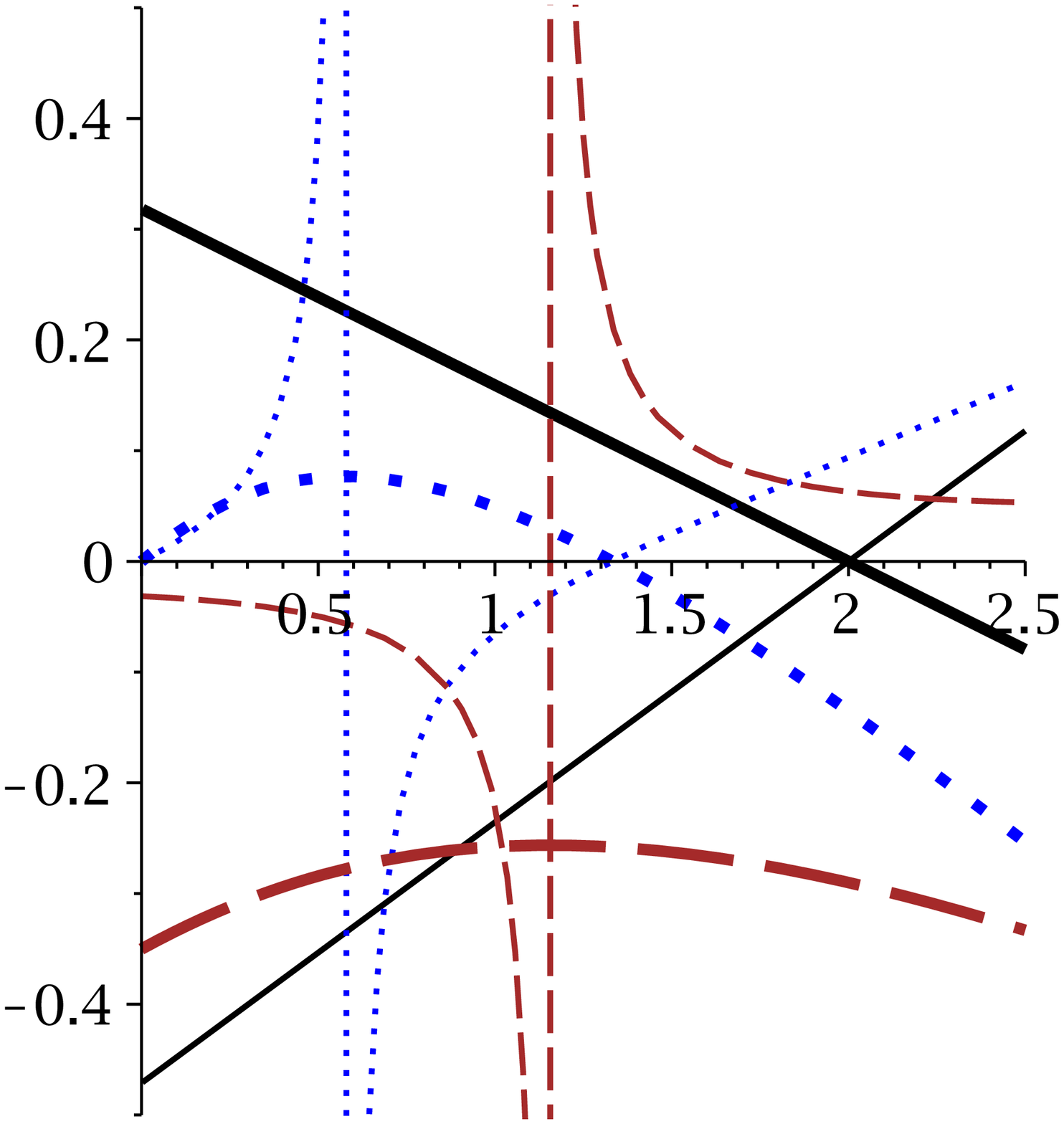} & \epsfxsize=6.5cm %
\epsffile{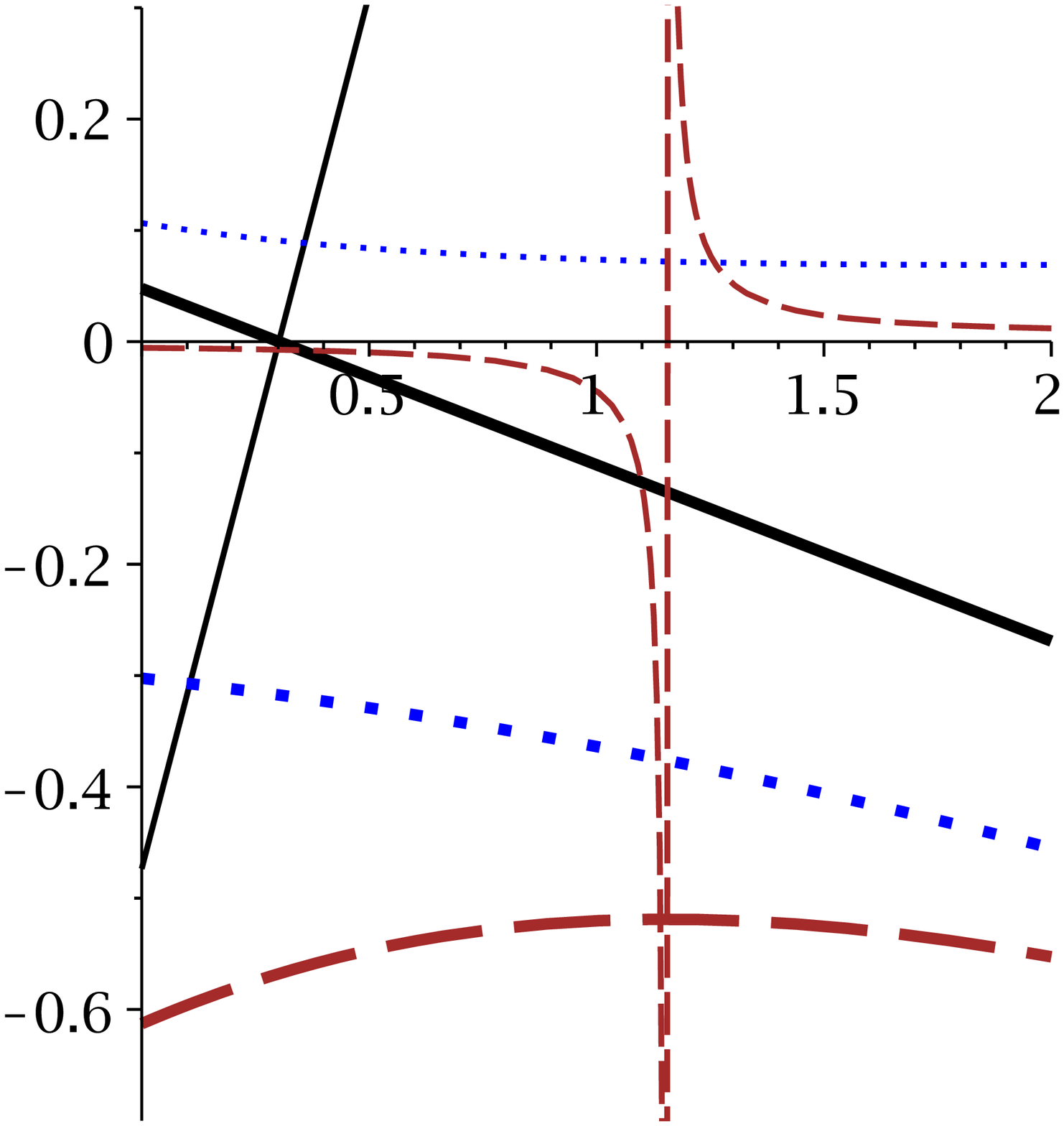}%
\end{array}
$%
\caption{For different scales: $C_{Q}$ and $T$ (bold lines) versus $r_{+}$
for $c=c_{1}=1$, $\Lambda=1$ and $m=2$. \newline
Left panel: $\protect\beta=1$, $q=0$ (continuous line), $q=1$ (dotted line)
and $q=2$ (dashed line). \newline
Right panel: $q=2$, $\protect\beta=0.05$ (continuous line), $\protect\beta%
=0.6$ (dotted line) and $\protect\beta=1$ (dashed line).}
\label{Fig33dS}
\end{figure}

It is evident that bound point is a function of cosmological constant,
massive and BI parameters. For adS black holes, only one bound point could
be observed while for dS black holes (by satisfying specific condition) two
bound points could be obtained. It should be pointed out that existence of
two bound points for dS configuration is due to contribution of the massive
gravity. In other words, in absence of massive gravity, only one bound point
is obtainable for these black holes in dS spacetime. Contrary to bound
point, the phase transition point is independent of the massive gravity. No
contribution of the massive gravity is observed for $r_{c}$. The dS black
holes enjoy the existence of second order phase transition in their phase
space, while such transition does not exist for adS black holes. The phase
transition point is a function of the electric charge, BI parameter and $%
\Lambda$. It is evident that generalization to BI nonlinear electrodynamics
has affected thermodynamical structure of the solutions (dependency of the
bound point and phase transition point on BI parameter). In order to
elaborate the effects of different parameters, we plot some diagrams for dS
and adS spacetimes (Figs. \ref{Fig2}, \ref{Fig3} for adS; Figs. \ref{Fig22dS}
and \ref{Fig33dS} for dS).

%%%%%%%%%%%%%%%%%%%%%%%%%%%%%%%%%%%%%%%%%%%%%%%%%%%%%%%%%%%%%%%

%%%%%%%%%%%%%%%%%%%%%%%%%%%%%%%%%%%%%%%%%%%%%%%%%%%%%%%%%%%%%%%
%%%%%%%%%%%%%%%%%%%%%%%%%%%%%%%%%%%%%%%%%%%%%%%%%%%%%%%%%%%%%%%

%%%%%%%%%%%%%%%%%%%%%%%%%%%%%%%%%%%%%%%%%%%%%%%%%%%%%%%%%%%%%%%

%%%%%%%%%%%%%%%%%%%%%%%%%%%%%%%%%%%%%%%%%%%%%%%%%%%%%%%%%%%%%%%

%%%%%%%%%%%%%%%%%%%%%%%%%%%%%%%%%%%%%%%%%%%%%%%%%%%%%%%%%%%%%%%
%%%%%%%%%%%%%%%%%%%%%%%%%%%%%%%%%%%%%%%%%%%%%%%%%%%%%%%%%%%%%%%

%%%%%%%%%%%%%%%%%%%%%%%%%%%%%%%%%%%%%%%%%%%%%%%%%%%%%%%%%%%%%%%

Plotted diagrams of adS solutions show that bound point is a decreasing
function of the massive parameter (left panel of Fig. \ref{Fig2}) and
cosmological constant (right panel of Fig. \ref{Fig2}), while it is an
increasing function of the electric charge (left panel of Fig. \ref{Fig3})
and BI parameter (right panel of Fig. \ref{Fig3}). For the absence of
electric charge, solutions are physical and stable for every horizon radius.
One of the interesting results of the plotted diagrams is regarding the
variation of BI parameter. For small values of this parameter, the behavior
of the temperature is Schwarzschild like \cite{HendiJHEP}. Here, similar to
the absence of electric charge, for sufficiently small values of
nonlinearity parameter, black holes are physical and thermally stable for
all the values of the horizon radius. In addition, similar to neutral case,
for the limit $r_{+}\longrightarrow 0$ (black holes evaporation), the
temperature and heat capacity are non-zero which has similar phenomenology
that was stated before with one important exception; in this case black
holes are not neutral. They are nonlinearly charged.

The case of dS solutions have more variations comparing to adS case. It is
evident that depending on choices of different parameters, the temperature
could be only a decreasing function of the horizon radius or it may acquire
a maximum (see Figs. \ref{Fig22dS} and \ref{Fig33dS} for more details). The
maximum of temperature is a decreasing function of the electric charge and
BI parameter (Fig. \ref{Fig33dS}), while it is an increasing function of the
massive gravity (Fig. \ref{Fig22dS} left panel). In place of maximum, if it
is positive, there is a phase transition of the larger to smaller black
holes.

For the case of temperature being a decreasing function of the horizon
radius, there may exist a bound point. If bound point exists, a region of
positive temperature exists. In physical region heat capacity is negative,
therefore, physical solutions are instable, vise versa is observed for the
absence of bound point. It is evident that for dS black holes, existence of
non-zero temperature for $r_{+}=0$ is achieved here much more easier
comparing to adS solutions. But in this case for $r_{+}=0$ while temperature
is positive, heat capacity is negative (Fig. \ref{Fig22dS} right panel).

The basic motivation for generalizing to nonlinearity is for solving
shortcomings of the Maxwell theory and introducing new phenomenology. We see
that generalization to nonlinearity leads to existence of the non-zero
temperature in case of charged black holes evaporation in the presence of
massive gravity. Existence of the such non-zero temperature highlights the
effect of the BI electromagnetic field. To summarize, one can state that
generalization to massive gravity leads to existence of non-zero temperature
in case of black holes evaporation for neutral black holes and
generalization to nonlinear electromagnetic field leads to existence of such
property for nonlinearly charged BTZ-massive black holes.

\begin{figure}[tbp]
$%
\begin{array}{c}
\epsfxsize=7cm \epsffile{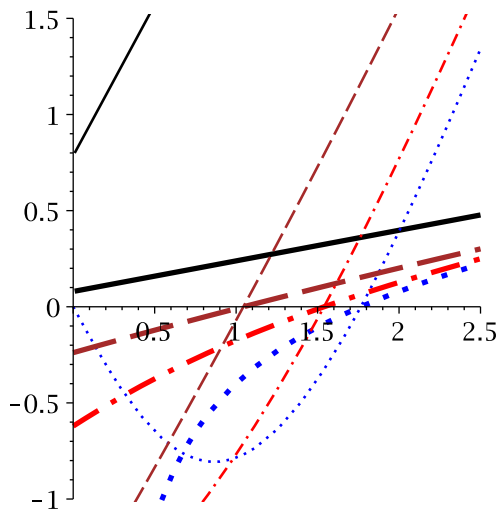}%
\end{array}
$%
\caption{$C_{Q}$ and $T$ (bold lines) versus $r_{+}$ for $c=c_{1}=1$, $%
\Lambda=-1$ and $m=1$. \newline
Neutral case (continuous line), Maxwell case for $q=1$ (dotted line), BI
case with $q=2$ for $\protect\beta=0.5$ (dashed line) and $\protect\beta=1.1$
(dashed-dotted line).}
\label{Fig4}
\end{figure}

Our final study in this section is making a simple comparison between
neutral, Maxwell and BI theories (see Fig. \ref{Fig4}). It is evident that
the largest bound point belongs to Maxwell theory (dotted line in Fig. \ref%
{Fig4}). No bound point is observed for neutral case (continuous line in
Fig. \ref{Fig4}) and the bound points of BI theory for different $\beta$ are
located between these two limits (dashed and dotted-dashed lines in Fig. \ref%
{Fig4}). These modifications highlight the effects of charged and
nonlinearly charged generalizations. To conclude, one can point out that the
physical stable black holes are formed in smaller horizon radius in BI
theory comparing to Maxwell one.
%%%%%%%%%%%%%%%%%%%%%%%%%%%%%%%%%%%%%%%%%%%%%%%%%%%%%%%%%%%%%%%

%%%%%%%%%%%%%%%%%%%%%%%%%%%%%%%%%%%%%%%%%%%%%%%%%%%%%%%%%%%%%%%

\section{Critical Behavior}

\label{Critical Behavior}

In this section, we will investigate the possible existence of Van der Waals
like behavior for obtained solutions by using the proportionality between
thermodynamical pressure and negative cosmological constant (adS solution).
It was shown that by considering cosmological constant as thermodynamical
pressure, it is possible to enrich thermodynamical structure of black holes
and introduce new phenomena such Van der Waals like phase transition. The
proportionality is given by
\begin{equation}
P=-\frac{\Lambda }{8\pi }.  \label{P}
\end{equation}

Here, instead of using usual procedure to study the possibility of the
critical behavior for these black holes, we employ another method which was
introduced in Ref. \cite{int}. This method is based on using the denominator
of the heat capacity. By replacing cosmological constant with its
corresponding pressure and solving the denominator of the heat capacity with
respect to pressure, one can obtain a relation for $P$. This relation is
different form equation of state. The maximums of the obtained relation for
pressure are places in which phase transition takes place and system has Van
der Waals like behavior. In other words, the pressure and horizon radius of
the maximum are critical pressure and horizon radius. This method was
employed in several paper \cite%
{HendiMassive1,HendiMassive2,HendiMassive4,HendiRainbow}. Now, we employ
this method to study critical behavior of the system.

Using the obtained heat capacities for Maxwell and BI cases (Eqs. (\ref%
{CQMax}) and (\ref{CQBI}), respectively) with Eq. (\ref{P}), one can find
following relations by solving the denominator of these heat capacities with
respect to pressure,
\begin{eqnarray}
P_{Maxwell} &=&-\frac{q^{2}}{8\pi r_{+}^{2}}, \\
&&  \notag \\
P_{BI} &=&-\frac{\beta ^{2}q^{2}}{4\pi \left( q^{2}+\beta
^{2}r_{+}^{2}\left( 1+\Gamma _{+}\right) \right) }.
\end{eqnarray}

Since electric charge and BI parameter are positive quantities, the obtained
relations indicate that no positive maximum for pressure exists for Maxwell
and BI cases. Therefore, even by considering the generalization of the
massive gravity, Van der Waals like behavior does not exists for charged
three dimensional massive black holes in presence of cosmological constant.
This result is expected. By taking a closer look at the heat capacity, one
can see that denominator of the heat capacity is independent of the massive
gravity. This independency results into absence of the Van der Waals like
behavior (previously, it was shown that BTZ black holes suffer the absence
of Van der Waals like behavior in their phase diagrams).

A question may rise regarding consideration of the positive cosmological
constant proportional to thermodynamical pressure. Calculations for this
case will lead to following equations for pressure
\begin{eqnarray}
P_{Maxwell} &=&\frac{q^{2}}{8\pi r_{+}^{2}}, \\
&&  \notag \\
P_{BI} &=&\frac{\beta ^{2}q^{2}}{4\pi \left( q^{2}+\beta ^{2}r_{+}^{2}\left(
1+\Gamma _{+}\right) \right) }.
\end{eqnarray}

The obtained relations for pressure confirm that there is no maximum for
pressure in this case. Therefore, following previously mentioned conditions
for existence of phase transition, one can conclude that no phase transition
exists for this case either.

\section{Geometrical thermodynamics}

\label{Geometrical thermodynamics}

Here, we regard the geometrical approach toward studying thermodynamical
structure of the solutions. In this approach, thermodynamical phase space of
a black hole is constructed by employing one of the thermodynamical
quantities as potential. Other extensive thermodynamical quantities are
components of the phase space. The thermodynamical information of the system
is encoded within Ricci scalar of this thermodynamical metric. The bound
points and phase transitions that were obtained in studying heat capacity,
are matched with divergencies of thermodynamical Ricci scalar of the
constructed phase space. In other words, divergencies of the Ricci scalar
are points in which system acquire bound points or meet phase transitions.
Since, this is another method for studying thermodynamical behavior of the
system, the results of it must be consistent with those obtained through
other methods. Meaning, that ensemble dependency must not be observed.

%%%%%%%%%%%%%%%%%%%%%%%%%%%%%%%%%%%%%%%%%%%%%%%%%%%%%%%%%%%%%%%
\begin{figure}[tbp]
$%
\begin{array}{cc}
\epsfxsize=6.5cm \epsffile{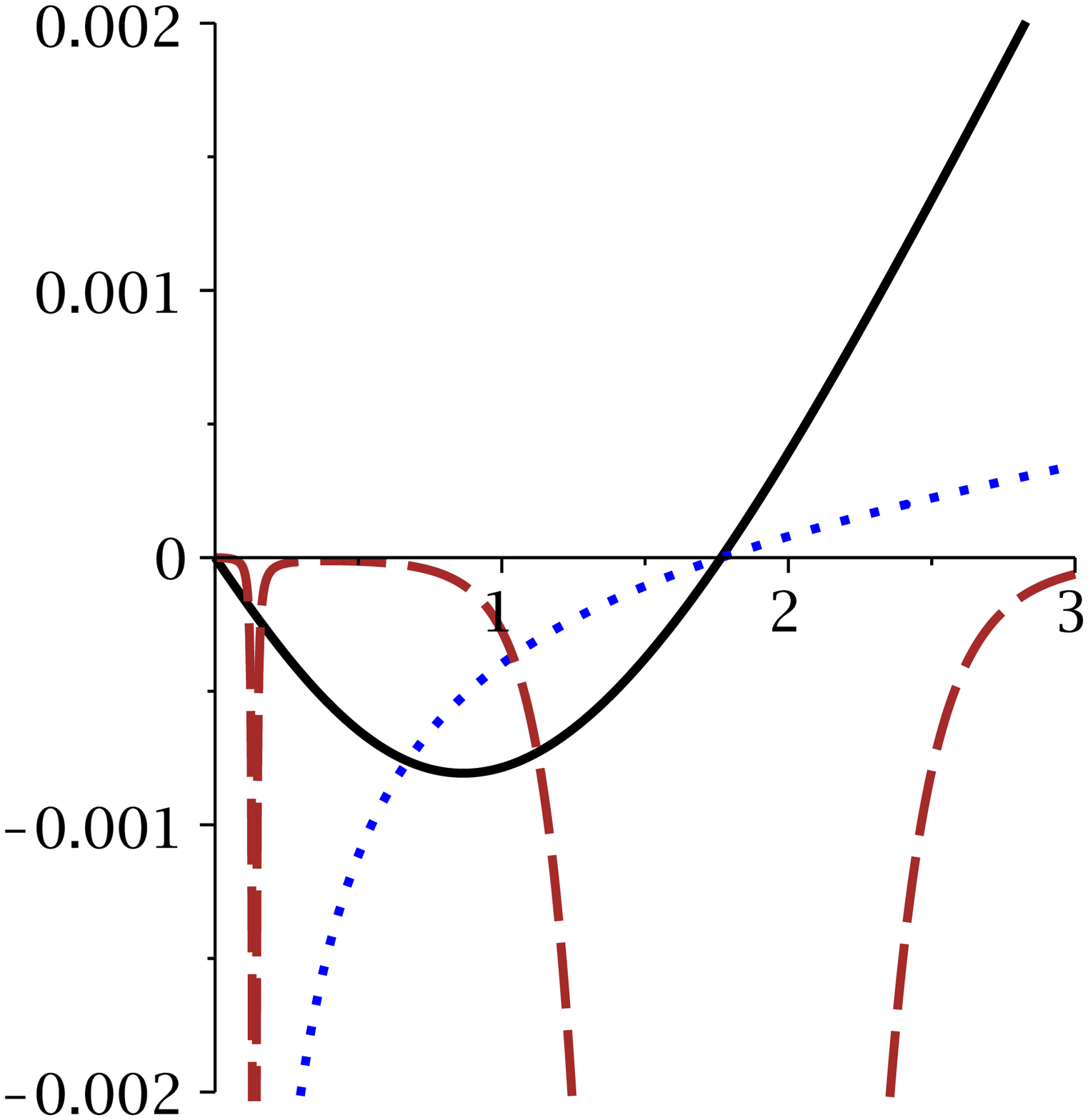} & \epsfxsize=6.5cm %
\epsffile{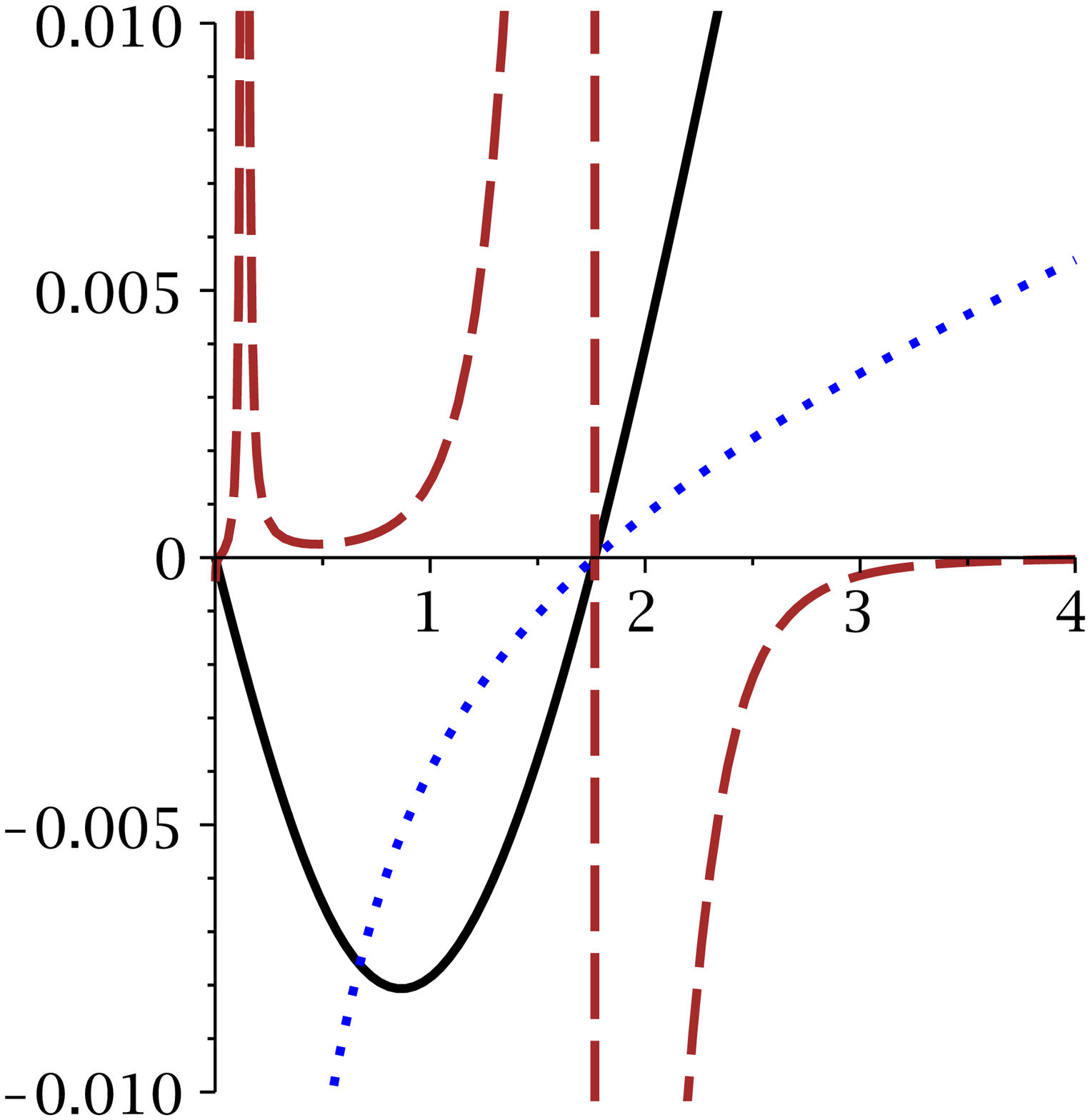}%
\end{array}
$%
\caption{For different scales: $C_{Q}$ (continuous line), $T$ (dotted line)
and $\mathcal{R}$ (dashed line) versus $r_{+}$ for $c=c_{1}=1$, $\Lambda=-1$%
, $q=2$ and $m=1$. \newline
Left panel: Weinhold, right panel: Ruppeiner "for Maxwell case".}
\label{Fig5}
\end{figure}

%%%%%%%%%%%%%%%%%%%%%%%%%%%%%%%%%%%%%%%%%%%%%%%%%%%%%%%%%%%%%%%
%%%%%%%%%%%%%%%%%%%%%%%%%%%%%%%%%%%%%%%%%%%%%%%%%%%%%%%%%%%%%%%
\begin{figure}[tbp]
$%
\begin{array}{ccc}
\epsfxsize=5.5cm \epsffile{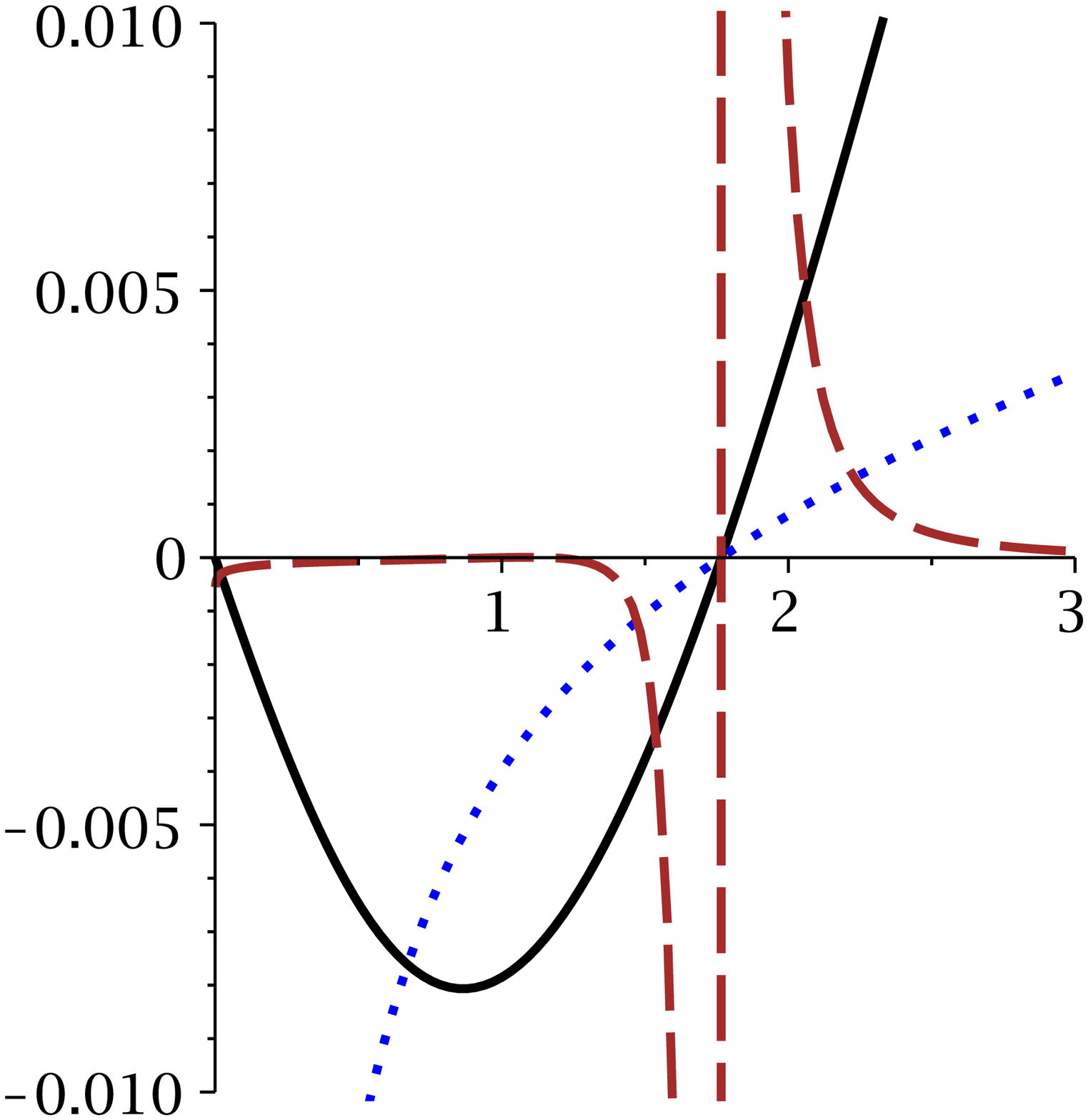} & \epsfxsize=5.5cm %
\epsffile{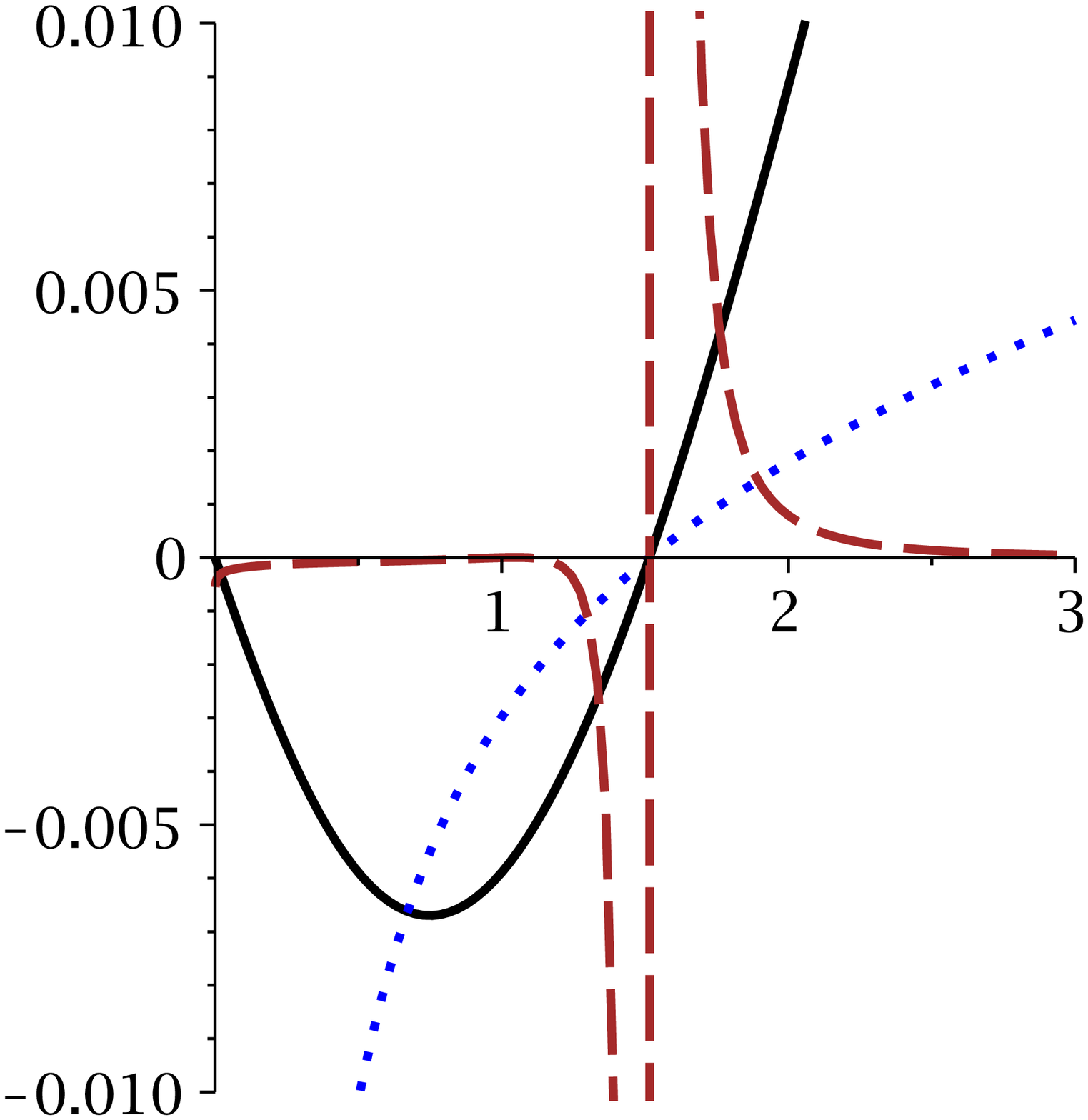} & \epsfxsize=5.5cm %
\epsffile{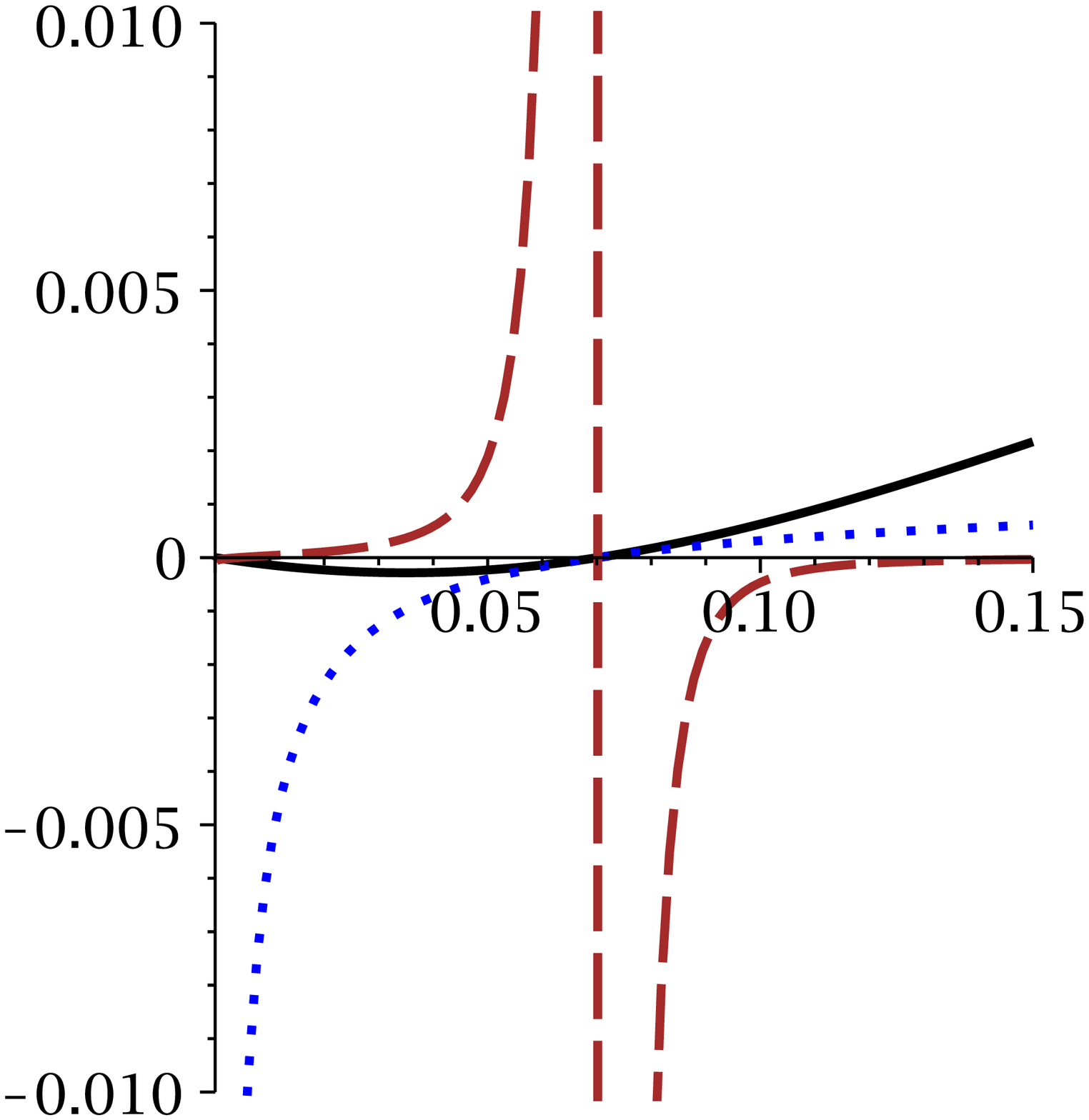}%
\end{array}
$%
\caption{For different scales: $C_{Q}$ (continuous line), $T$ (dotted line)
and $\mathcal{R}$ (dashed line) versus $r_{+}$ for $c=c_{1}=1$ and $%
\Lambda=-1$. \newline
Left panel: $q=2$ and $m=1$, middle panel: $q=2$ and $m=1.5$, right panel: $%
q=0.2$ and $m=1$. \newline
HPEM metric "for Maxwell case".}
\label{Fig6}
\end{figure}

%%%%%%%%%%%%%%%%%%%%%%%%%%%%%%%%%%%%%%%%%%%%%%%%%%%%%%%%%%%%%%%
%%%%%%%%%%%%%%%%%%%%%%%%%%%%%%%%%%%%%%%%%%%%%%%%%%%%%%%%%%%%%%%
\begin{figure}[tbp]
$%
\begin{array}{cc}
\epsfxsize=6.5cm \epsffile{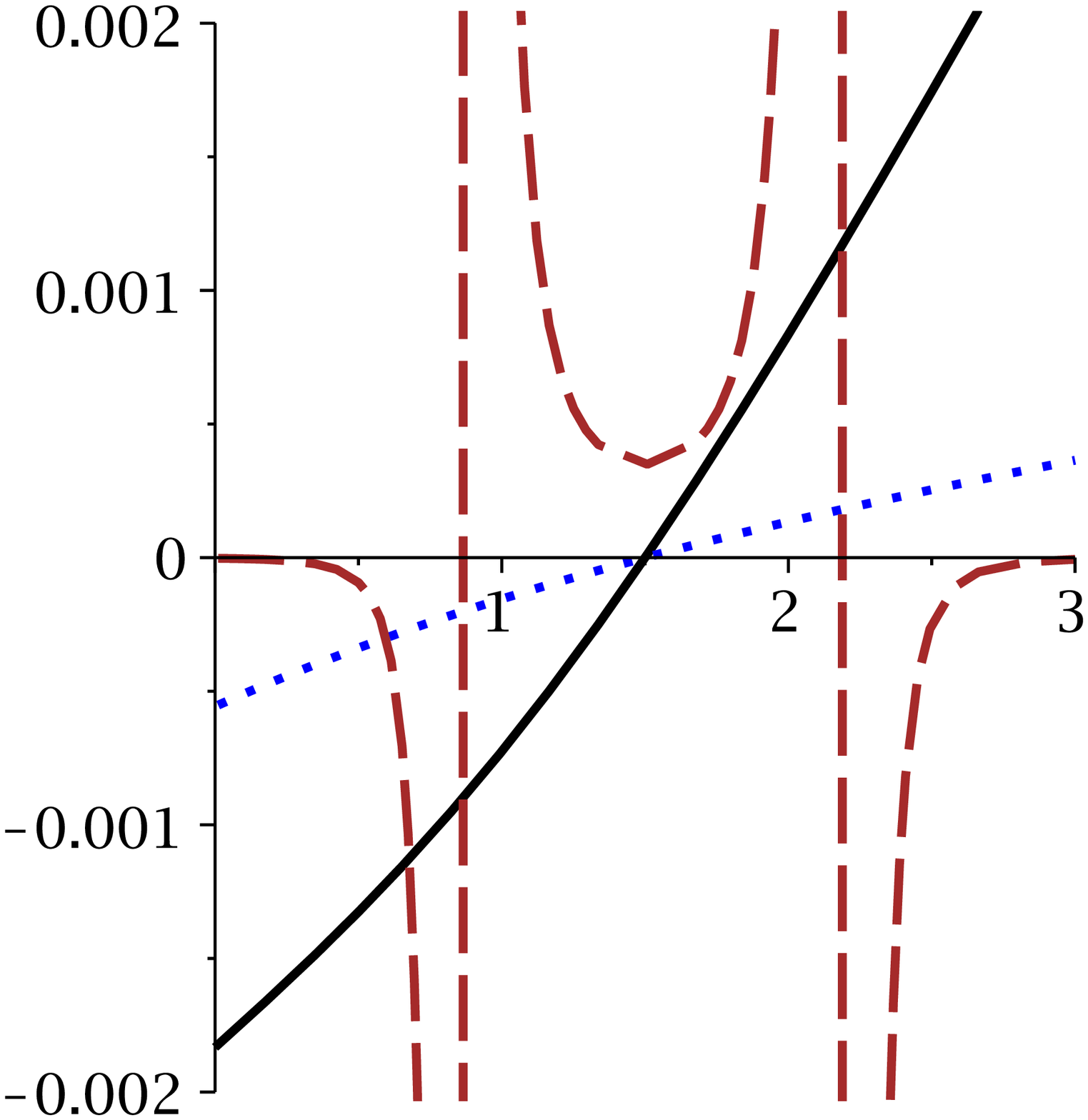} & \epsfxsize=6.5cm %
\epsffile{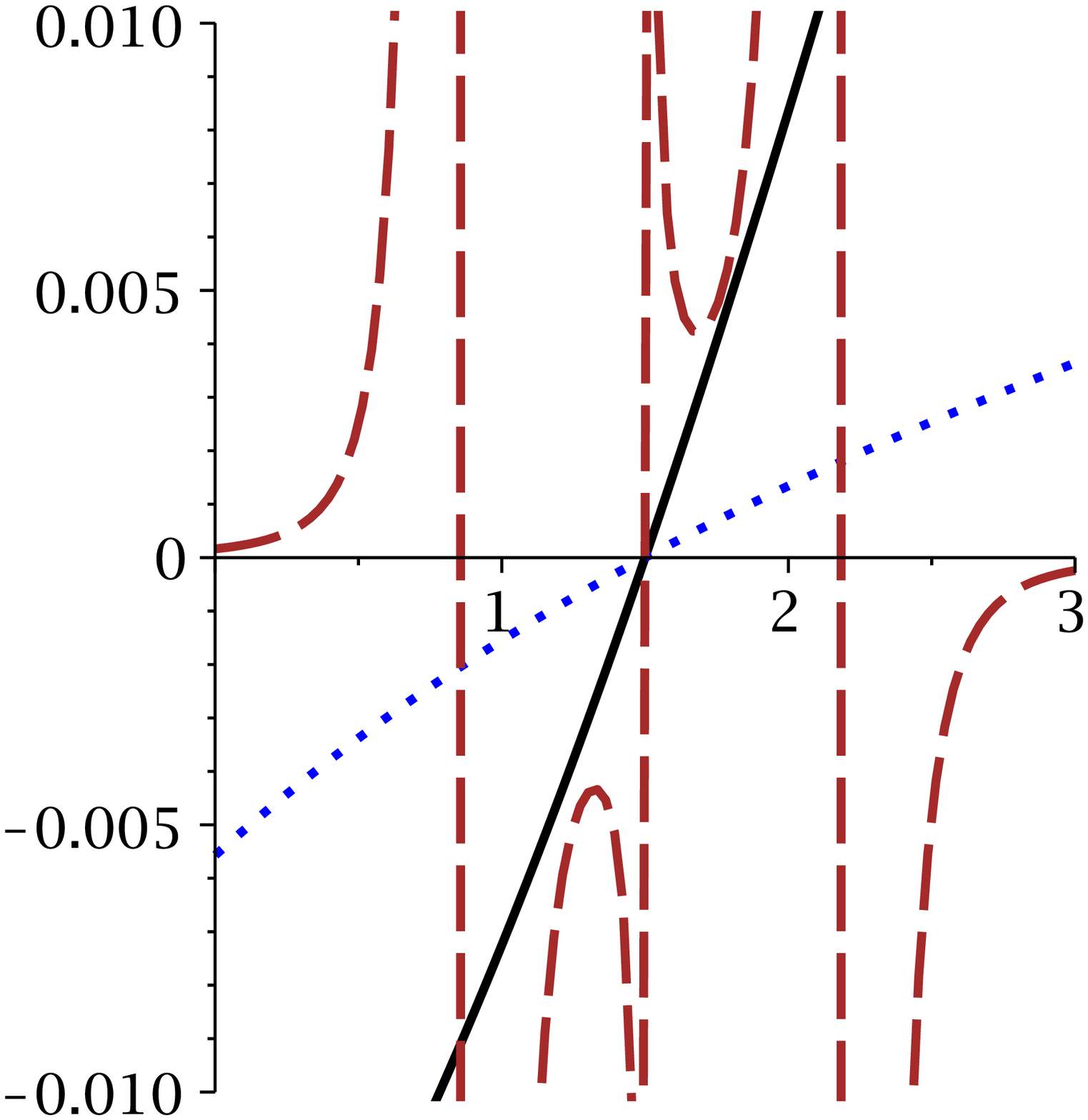}%
\end{array}
$%
\caption{For different scales: $C_{Q}$ (continuous line), $T$ (dotted line)
and $\mathcal{R}$ (dashed line) versus $r_{+}$ for $c=c_{1}=1$, $\Lambda=-1$%
, $q=2$, $\protect\beta=1$ and $m=1$. \newline
Left panel: Weinhold, right panel: Ruppeiner "for BI case".}
\label{Fig7}
\end{figure}

%%%%%%%%%%%%%%%%%%%%%%%%%%%%%%%%%%%%%%%%%%%%%%%%%%%%%%%%%%%%%%%
%%%%%%%%%%%%%%%%%%%%%%%%%%%%%%%%%%%%%%%%%%%%%%%%%%%%%%%%%%%%%%%
\begin{figure}[tbp]
$%
\begin{array}{ccc}
\epsfxsize=5.5cm \epsffile{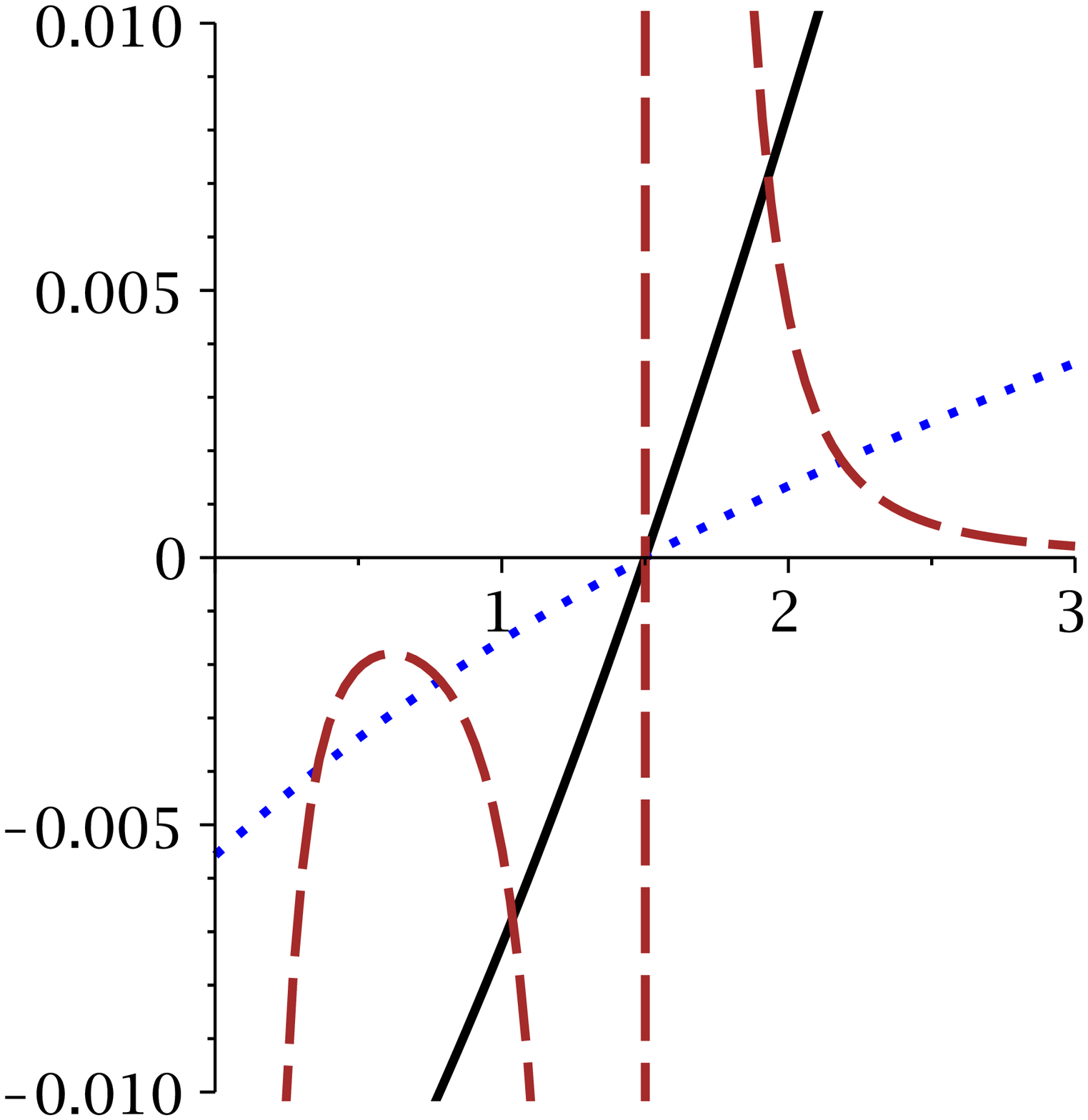} & \epsfxsize=5.5cm %
\epsffile{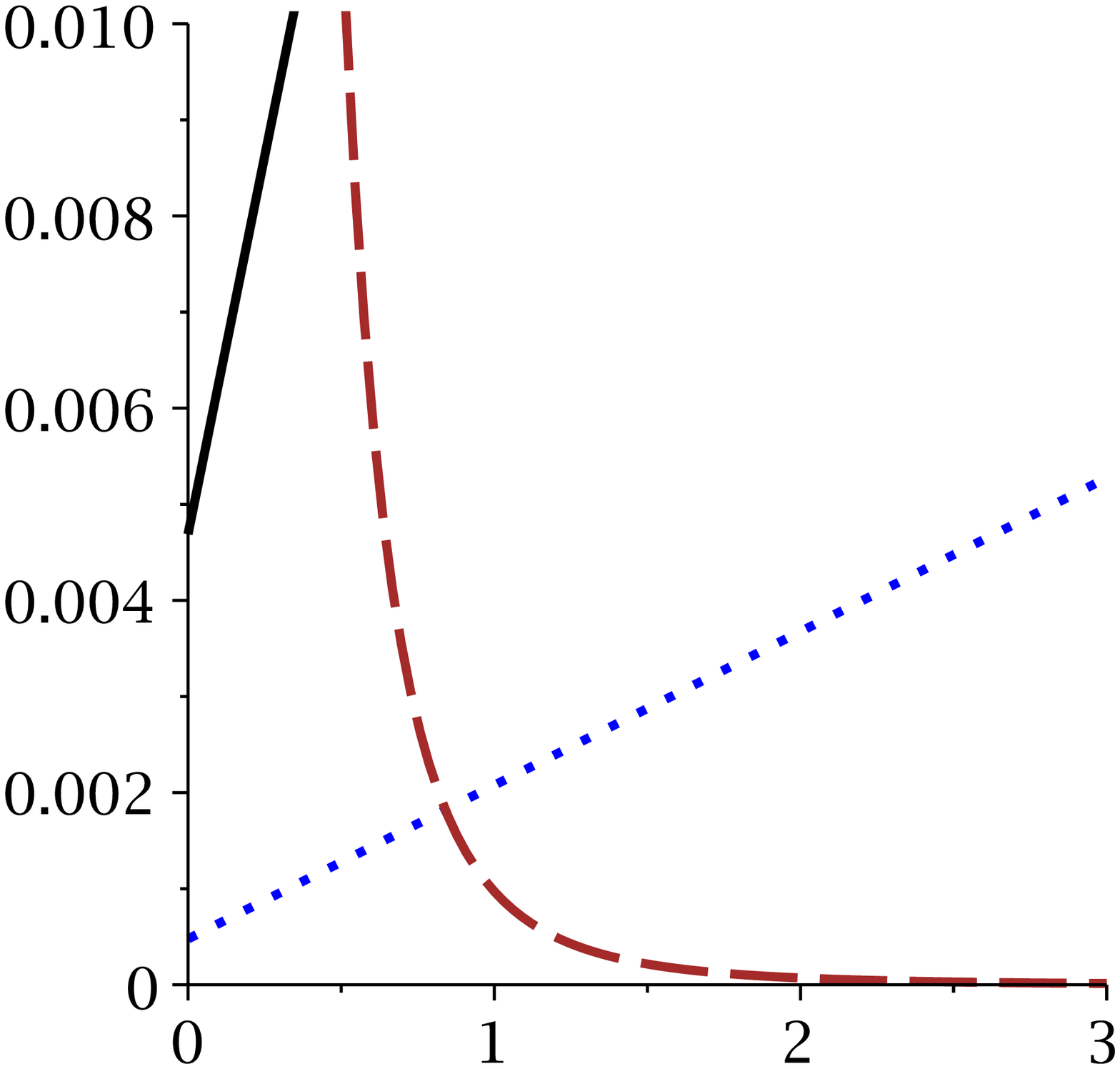} & \epsfxsize=5.5cm \epsffile{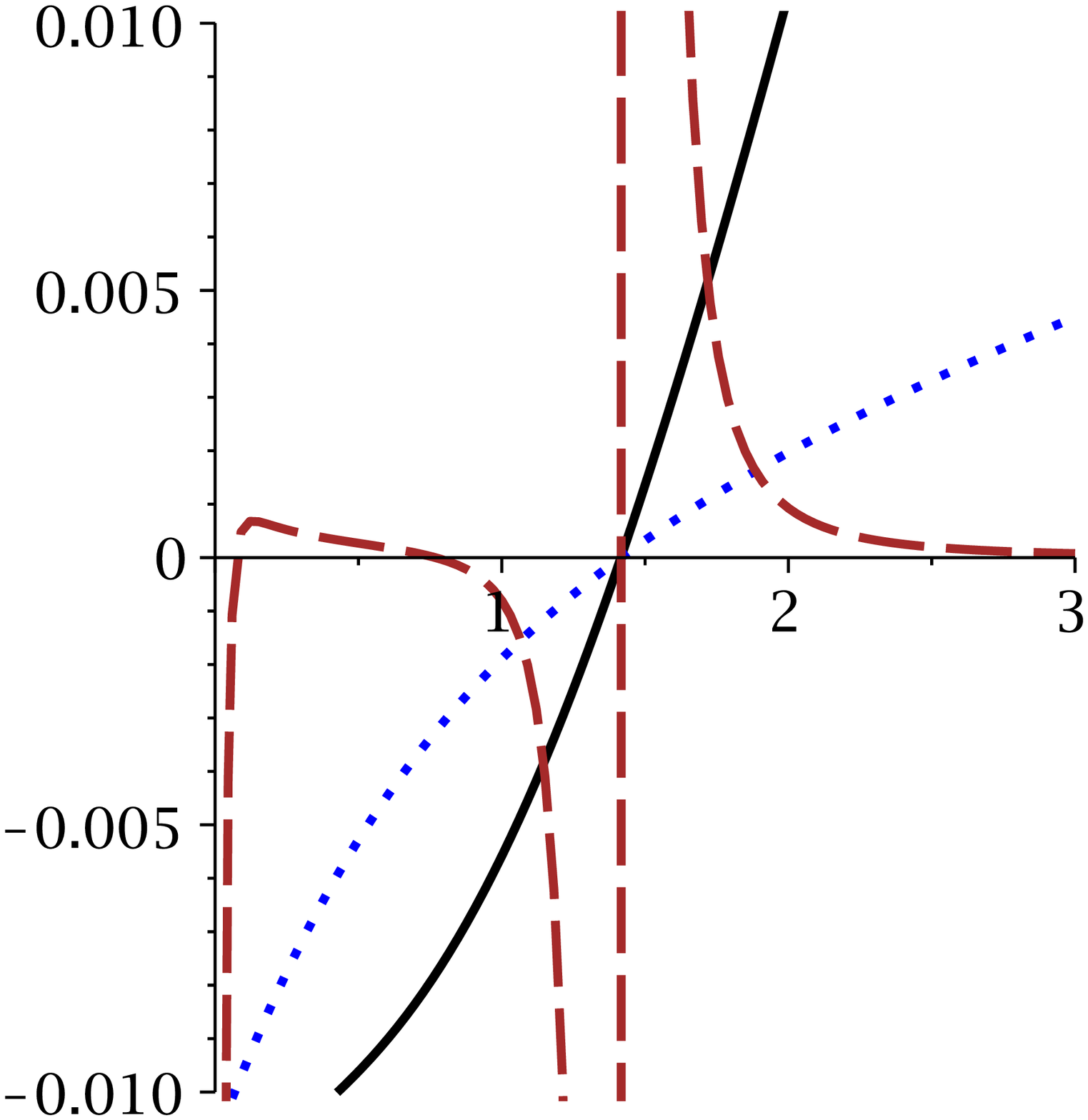}%
\end{array}
$%
\caption{For different scales: $C_{Q}$ (continuous line), $T$ (dotted line)
and $\mathcal{R}$ (dashed line) versus $r_{+}$ for $c=c_{1}=1$, $q=2$ and $%
\Lambda=-1$. \newline
Left panel: $\protect\beta=1$ and $m=1$, middle panel: $\protect\beta=0.05$
and $m=1$, right panel: $\protect\beta=2$ and $m=1.5$. \newline
HPEM metric "for BI case".}
\label{Fig8}
\end{figure}

%%%%%%%%%%%%%%%%%%%%%%%%%%%%%%%%%%%%%%%%%%%%%%%%%%%%%%%%%%%%%%%

A successful method of geometrical thermodynamics covers all bound and phase
transition points that were observed in heat capacity. In other words, no
mismatch between divergencies of the Ricci scalar and mentioned points
should be observed and no extra divergency should exist. There are several
methods toward geometrical thermodynamics. The well known ones are: Weinhold
\cite{WeinholdI,WeinholdII}, Ruppeiner \cite{RuppeinerI,RuppeinerII},
Quevedo \cite{QuevedoI,QuevedoII} and HPEM \cite{HPEMI,HPEMII,HPEMIII}.
Their corresponding metrics are in following forms
\begin{equation}
ds^{2}=\left\{
\begin{array}{cc}
Mg_{ab}^{W}dX^{a}dX^{b} & Weinhold \\
&  \\
-T^{-1}Mg_{ab}^{R}dX^{a}dX^{b} & Ruppeiner \\
&  \\
\left( SM_{S}+QM_{Q}\right) \left( -M_{SS}dS^{2}+M_{QQ}dQ^{2}\right)  &
Quevedo\;\;\;Case\;\;I \\
&  \\
SM_{S}\left( -M_{SS}dS^{2}+M_{QQ}dQ^{2}\right)  & Quevedo\;\;\;\;Case\;\;II
\\
&  \\
S\frac{M_{S}}{M_{QQ}^{3}}\left( -M_{SS}dS^{2}+M_{QQ}dQ^{2}\right)  & HPEM%
\end{array}%
\right. ,
\end{equation}%
where $M_{X}=\partial M/\partial X$ and $M_{XX}=\partial ^{2}M/\partial X^{2}
$. Using these metrics, it was shown that denominators of Ricci scalar of
these phase spaces are \cite{HPEMI}%
\begin{equation}
denom(\mathcal{R})=\left\{
\begin{array}{cc}
M^{2}\left( M_{SS}M_{QQ}-M_{SQ}^{2}\right) ^{2} & Weinhold \\
&  \\
M^{2}T\left( M_{SS}M_{QQ}-M_{SQ}^{2}\right) ^{2} & Ruppeiner \\
&  \\
M_{SS}^{2}M_{QQ}^{2}\left( SM_{S}+QM_{Q}\right) ^{3} & Quevedo\;I \\
&  \\
2S^{3}M_{SS}^{2}M_{QQ}^{2}M_{S}^{3} & Quevedo\;II \\
&  \\
2S^{3}M_{SS}^{2}M_{S}^{3} & HPEM%
\end{array}%
\right. .
\end{equation}

The factors $M_{SS}$ and $M_{S}$ in denominators of the HPEM and Quevedo
metrics ensure that divergencies of the Ricci scalars are matched with bound
and phase transition points, whereas, the factor $M_{QQ}$ in Quevedo metrics
provides an extra divergency for Ricci scalar of these metrics which is not
matched with any corresponding point in the heat capacity. Analytical
calculations show that for Maxwell and BI cases of this paper, one can find
following roots for the factor $M_{QQ}$
\begin{eqnarray}
r_{0-Maxwell} &=&4\ln \left( \frac{r_{+}}{l}\right) ^{2}, \\
&&  \notag \\
r_{0-BI} &=&\frac{4\left[ \left( 2r_{+}^{2}\beta ^{2}\left( 1-\Gamma
_{+}\right) -q^{2}\right) \Gamma _{+}\ln \left( \frac{\left( 1+\Gamma
_{+}\right) \pi r_{+}}{2l}\right) +\left( 1+\Gamma _{+}\right) \left(
2r_{+}^{2}\beta ^{2}\ln \pi -q^{2}\right) +q^{2}\left( 2+\Gamma _{+}\right)
\ln \pi \right] ^{2}}{r_{+}^{4}\beta ^{4}\Gamma _{+}^{2}\left( 1+\Gamma
_{+}\right) ^{4}},
\end{eqnarray}
which means that for Quevedo metrics, at least, one extra divergency for
their Ricci scalar is observed that is not coincidence with any bound and
phase transition points. Therefore, employing Quevedo metrics results into a
case of ensemble dependency. As for Ruppeiner and Weinhold, it was not
possible to obtain divergencies of their Ricci scalar analytically.
Therefore, we use numerical evaluation and present results in some diagrams
(see Figs. \ref{Fig5}-\ref{Fig7}).

It is evident that divergencies of the Ricci scalar of Weinhold (left panels
of Figs \ref{Fig5} and \ref{Fig7}) and Ruppeiner (right panels of Figs \ref%
{Fig5} and \ref{Fig7}) are not matched completely with bound and phase
transition points. This result indicates that using these thermodynamical
metrics also leads to ensemble dependency. On the contrary, the HPEM metric
provides divergencies in its Ricci scalar which are coincidence with bound
and phase transition points (see Figs \ref{Fig6} and \ref{Fig8}). In other
words, no extra divergency and mismatched are observed. Therefore, using
this method results into a uniform picture regarding bound points and phase
transitions. In other words, no ensemble dependency exists for this method.

\section{Conclusions}

In this paper, we have considered (non)linearly charged BTZ black holes in
the massive gravity context. After investigating geometrical properties,
thermodynamical structure of the solutions has been investigated through
various methods. It was shown that thermodynamical structure of the
solutions is highly sensitive to consideration of the adS or dS spacetime.
The number of bound points and phase transitions was a function of
cosmological constant and was sensitive to the sign of $\Lambda$.

Interestingly, it was shown that due to contribution of the massive gravity,
number of the bound points was modified while the divergencies (phase
transition) of the heat capacity was independence of the massive gravity.
One of the most important results of this paper was existence of non-zero
temperature for vanishing horizon radius (final state of black holes after
evaporation). In other words, in case of neutral black holes, for black
holes evaporation, the temperature and heat capacity were non-zero,
indicating that after black holes evaporation, all information of the black
holes is not lost. A trace of existence of black holes is left in terms of a
fluctuation of the temperature. Such property was not observed for linearly
charged black holes, whereas, interestingly, generalization to nonlinear
electromagnetic field recovered this property. It means that after
evaporation of the nonlinearly charged BTZ-massive black holes, the trace of
their existence will remain (non-zero temperature). In addition, it was
shown that generalization to nonlinearity also modified the place of bound
points, hence the formation of the physical black holes.

Next, it was shown that despite the generalization to massive gravity and
nonlinear electromagnetic field, these black holes suffer the absence of Van
der Waals like behavior in their phase diagrams.

In addition, geometrical methods was employed to study the thermodynamical
structure of the solutions. It was analytically shown that due to structure
of the Quevedo metrics, employing these metrics, leads to existence of
ensemble dependency. In other words, the bound points and phase transitions
were not matched with divergencies of the Ricci scalar of Quevedo metrics,
completely and/or extra divergencies were observed. In addition, it was
pointed out that Weinhold and Ruppeiner metrics also leads to presence of
ensemble dependency. Whereas, the HPEM method, was free of mentioned
problems and provided a uniform picture for studying thermodynamical
structure of the solutions.

\begin{acknowledgements}

We thank Shiraz University Research Council. This work has been
supported financially by the Research Institute for Astronomy and
Astrophysics of Maragha, Iran.
\end{acknowledgements}

\end{document}